\documentclass[useAMS,usenatbib]{mnras}
\usepackage[latin1]{inputenc}
\usepackage[english]{babel}
\usepackage{graphicx}
\usepackage{amsmath}
\usepackage{amsfonts}
\title[Cosmic Variation of Proton to Electron Mass Ratio with an interacting Higgs Scalar Field]{Cosmic Variation of Proton to Electron Mass Ratio with an interacting Higgs Scalar Field}
\author[Soumya Chakrabarti]{Soumya Chakrabarti\thanks{E-mail : soumya.chakrabarti@saha.ac.in}\\
Theory Division\\
Saha Institute of Nuclear Physics\\
Kolkata 700064\\
India}
\date{Accepted XXX. Received YYY; in original form ZZZ}
\pubyear{2021}
\begin{document}
\maketitle
\begin{abstract}
We discuss that it is quite possible to realize the smooth transition of the universe between a matter/radiation dominated deceleration and a dark energy dominated acceleration, even with a variation of proton-to-electron mass ratio $\mu$. The variation is incorporated into the theory of gravity using a cosmological Higgs scalar field with a non-trivial self-interaction potential, leading to a varying Higgs vacuum expectation value (VEV). This matches well with the data from molecular absorption spectra of a series of Quasars. In comparison with late-time cosmology, an observational consistency is reached using a Markov chain Monte Carlo simulation and JLA+OHD+BAO data sets. We find that the pattern of variation is embedded within the evolving Equation of State (EOS) of the scalar Dark Energy/Matter components, but leaves negligible trace on the effective EOS of the system. We discuss three cases of scalar extended theory of gravity, (a) a minimally coupled scalar, (b) a non-minimally coupled scalar and (c) a generalized Brans-Dicke setup. We also give a toy model of a unified cosmic history from inflation to the present era and discuss how the Higg VEV might have changed as a function of look back time.  
\end{abstract}
\begin{keywords}
cosmology: theory; dark energy; variation of fundamental constants
\end{keywords}
\section{Introduction}\label{s0}
Fundamental forces of the nature often elude our usual logical inferences. Gravity is one of these forces whose description requires curious mathematical notions. These notions collectively develop the language of General Theory of Relativity (GR). It has served as the best theory of gravity till date and succeeded in answering most of the puzzles coming from astrophysical observations. There are offcourse some inconclusive areas which leads to an eventual consideration of extended versions of GR. The most well-known amongst quite a few issues is the accelerated expansion of our Universe. The simplest way to explain this counter-intuitive phenomena is by considering a correction to the energy content of the universe, for example, of the order of a cosmological constant \citep{riess, eisen}. However, abysmal contradiction between the observational and the theoretical expectation of the energy scale makes it redundant. Apart from this, we also encounter a vacuum energy density of the same order of magnitude as the matter density, known as the famous coincidence problem. A time evolving scalar field, primarily motivated by fundamental theories of unification, often works as a prefatory entity of these issues altogether \citep{copeland}. It also serves a phenomenological purpose in the time history of the universe, as a candidate of interest in the epochs of present acceleration as well as the preceding deceleration \citep{paddy, riess0}. For a simple minimally interacting scalar this is initiated by the self-interaction of the field which stays dormant during deceleration and dominates during the late times.  \\

It is an essential problem in gravitational physics to properly characterize the dynamics of dark energy or the so-called fluid responsible for cosmic acceleration. Quite a few efforts have been made in this regard to write it's equation of state as a function of redshift, analytically as well as using observational comparisons such as the luminosity distance measurement of type Ia supernovae or weak lensing. Irrespective of these attempts, there is no clear idea about the distribution of dark energy. It does not cluster below Hubble scale but this does not give adequate information cosmologically. Moreover, the usual methodologies provide insufficient conclusions \citep{maor1, maor2}, which can be realized for instance, from a scalar field model of slow-roll dark energy \citep{slepian, anjansen}. These models effectively give a constant equation of state (EOS), disfavored in light of the luminosity distance measurement data \citep{paddy}. For a review of dark energy and it's EOS, different possibilities and future extensions, we refer to the work of \cite{upadhye}. A conclusive evidence of an evolving dark energy EOS can be difficult to come by from direct observations, but the requirement of a better theoretical understanding of the same is undeniable. In view of this, we focus on a philosophy that a time-evolving scalar field in the action of gravity can lead to the variation of standard constant couplings of the theory \citep{wetterich, carroll}. Nature of these couplings are paramount in the structure of standard cosmology and a variation surely generates distinguishable features. An idea of this class first came into being through a varying gravitational constant $G$ almost a century back and has since then received a plethora of theoretical and phenomenological treatments \citep{milne, dirac1, dirac2}. If the readers are interested in the relevance of varying natural constants we refer them to the reviews of \cite{uzan1, uzan2, chiba, calmet}. Phenomenological analysis of these theories are quite popular and often produce consequences of considerable interest \citep{livio, landau, chamoun, sandvik, olive, anchor, copeland1, bento, avelino, lee}. For instance, a theory with varying fine-structure constant $\alpha$ has received much attention \citep{parkinson, nunes, doran} and has shown observational validation in low redshift \citep{webb, murphy, chand, uzan1}. This has far reaching implications as in a time evolving scalar field theory standard gauge as well as the Yukawa couplings will vary alongwith a varying $\alpha$ \citep{campbell}. In the unified theories of particle physics a varying $\alpha$ also has a correlation with the strong interaction of quarks and gluons, through the QCD scale $\Lambda_{QCD}$ \citep{gasser}. On this note we go with the postulate that cosmologically, Higgs vacuum expectation value (VEV) can vary from it's present value (the VEV can be roughly interpreted as the difference between the origin and the minimum of a Higgs potential), for instance during the electroweak phase transition. In compliance with such a variation of Higgs VEV, the masses of the quarks are expected to vary. On the contrary, the proton mass is mainly determined by the QCD scale. Therefore, a varying Higgs VEV can generate variations in the proton-to-electron mass ratio as well as the mass scales of fundamental particles \citep{campbell, calmet0, lang, olive0, dine}. Defined as $\mu \equiv \frac{m_{p}}{m_{e}}$, this is a dimensionless ratio that quantifies strength of strong and electroweak scales. Typically a $\mu$-variation is related to fine structure constant through the relation 
\begin{equation}
\frac{\Delta\mu}{\mu} \sim \frac{\Delta\Lambda_{QCD}}{\Lambda_{QCD}} - \frac{\Delta{\nu}}{\nu} 
\sim R\frac{\Delta \alpha}{\alpha}. \label{defR}
\end{equation}
Typically $R$ is a large negative number (for instance, $R \sim -50$ as proved by \cite{avelino0}) and connected to high-energy scales in theories of unification. It dictates the analytical correlation between observations of a varying $\mu$ and a varying $\alpha$, however, this is entirely model-dependent. This is also supported by some limited yet intriguing observational data from molecular absorption spectra \citep{thompson, ivan, rien, flambaum, king, bagdon, rahmani, dapra}. They motivate one to look for evidences of a varying $\mu$ on a cosmological scale. Particularly in the context of a scalar extended theory of gravity, this motivation can lead one towards a better understanding of the cosmological constraints on the theory, nature of the scalar interaction(s) and the effective equation of state of the system. \\

In this manuscript we show that it is quite possible to construct an observationally viable late time cosmology using an interacting scalar field that allows a varying Higgs VEV. Similar motivations have led to non-trivial generalizations of GR in the recent past \citep{barrow, calmet1, dent, chiba0, lee0, avelino1, fritzsch}. For example, it has rejuvinated interest in a generalized Brans-Dicke theory, as it has been proved that with a Higgs scalar field with derivative interaction these theories can support an evolving Higgs VEV and the cosmic acceleration \citep{sola0, chakrabarti}. We generalize this and see if it is possible to conceive an evolving Higgs VEV and support a viable deceleration-to-acceleration transition with even the simplest possible scalar field models of cosmology. In particular we study the role of scalar interaction terms in the action and the evolving EOS parameters in the system. We also give a schematic idea regarding how the Higg VEV might have evolved in a unified cosmic history as a function of look back time. For most parts of the manuscript we rely on a recently demonstrated \citep{chakrabarti} method of cosmic reconstruction using the cosmic statefinder parameter. The next section is the `notion' part of the manuscript where we introduce our construct of a varying Proton-to-Electron Mass Ratio in a cosmological context. Section $3$ is the `execution' part with four subsections, where we adjudge the conjecture made in Section $2$ for different generalizations of scalar extended theory of gravity, before concluding in Section $4$.

\section{A Varying Proton-to-Electron Mass Ratio : The Origin and the Conjecture}
It is possible to realize a notion of varying proton-to-electron mass ratio in the theory of electroweak interactions, which forms a unification of electricity, magnetism and light. In this theory, a Higgs boson is the distributor of mass as the elementary particles have a mass proportional to $\nu$ or the Higgs VEV.  For any electron or quark, assuming the Yukawa coupling to be $\lambda_{e,q}$ one can write
\begin{equation}
m_{e,q} = \lambda_{e,q} \nu. \label{2-1}
\end{equation}
On the other hand, the proton mass $m_{p}$ receives most contribution from quark masses and the quark-gluon interaction, but has negligible effect from Higgs VEV (it can be proved from the quark mass expansion of $m_{p}$ and the separation of QCD Hamiltonian in gauge-invariant parts), leading to the following relation
\begin{equation}
m_{p} = a(\Lambda_{QCD}) + \sum\limits_{q} b_{q} m_{q}. \label{2-2}
\end{equation}
If $\Lambda_{QCD}$ (QCD energy scale) and the Yukawa coupling both are assumed to be constants \citep{calmet00}, from Eqs. (\ref{2-1}) and (\ref{2-2}) one can write
\begin{align}
\frac{\Delta m_{e}}{m_{e}} &= \frac{\Delta \nu}{\nu}, \label{2-8} \\
\frac{\Delta m_{p}}{m_{p}} &= \frac{\sum\limits_{q} b_{q} m_{q}}{a(\Lambda_{QCD}) + \sum\limits_{q} b_{q} m_{q}} \frac{\Delta \nu}{\nu} = \frac{9}{100} \frac{\Delta \nu}{\nu} . \label{2-9}
\end{align}
From Lattice QCD simulations \citep{gasser} it can be proved that the quark mass term $\sum\limits_{q} b_{q} m_{q}$ is accountable for less than $10$ percent of the proton mass. Thus a variation of Higgs VEV alone results in a negligible change in proton mass $\frac{\Delta m_{p}}{m_{p}}$ compared to $\frac{\Delta m_{e}}{m_{e}}$ \citep{ji, yang}. As an extension, we think it is also fair to assume that the proton mass does not vary much with cosmic time. On the other hand we take that cosmologically, the Higgs VEV sees atleast some mild variation from it's present value, taking into account the possibilities during an era of electroweak phase transitions. Therefore a varying Higgs VEV effectively leads to a varying proton-to-electron mass ratio. From Eqs. (\ref{2-8}) and (\ref{2-9})  
\begin{equation}
\frac{\Delta \mu}{\mu} = \frac{\Delta m_{p}}{m_{p}} - \frac{\Delta m_{e}}{m_{e}} =  - \frac{91}{100} \frac{\Delta \nu}{\nu}. \label{2-10}
\end{equation}
If $\nu_{0}$ and $\nu_{z}$ are the values of Higgs VEV at the present epoch (the present value of Higgs VEV is $\nu_{0} = 246 GeV$) and at some redshift $z$, then $\Delta \nu / \nu$ is equivalent to $(\nu_{z}-\nu_{0}) / \nu_{0}$. Apart from the unconventional theoretical considerations in the context of standard model particle physics, it is now known that strong gravitational fields can contribute to $\frac{\Delta \mu}{\mu}$ and subsequently affect cosmological effects in terms of look-back time \citep{bagdon0}.\\

To work with cosmological solutions with a varying Higgs VEV we first need a classical Higgs scalar field with an appropriate description. We write the standard model Higgs $\varphi$ (whose action is written as $L =  - \frac{1}{2} \partial_{\mu} \varphi  \partial^{\mu} \varphi - \frac{\lambda}{4} (\varphi^{2} -\nu^{2})^{2}$) as a classical background field $\phi$ and the Higgs particle $h$ \citep{calmet00, ahmed}
\begin{equation}
\varphi = \nu_{0} + \phi(t) + h = \nu(t) + h. 
\end{equation}
$\nu(t)$ is an evolving Higgs VEV and the classical field $\phi$ fits in as the cosmological Higgs. The manner in which this field $\phi$ interacts with gravity is a vital point as it leads to deviations in the field equations of the theory. Such a scalar has been used to generate a time evolution of $\nu$ in the recent past. For instance, it was discussed that a time-dependent Higgs VEV can generate non-adiabatic quantum effects leading to boson and fermion production \citep{casadio}. Also, in a scalar extended Brans-Dicke theory a Higgs scalar with slowly varying Higgs VEV is a necessary requirement in order to support power law cosmologies \citep{sola0}. In the present work, we assume a Higgs scalar with a mildly evolving VEV at the outset. We investigate a possible minimal coupling, a non-minimal coupling and non-minimal derivative coupling of the Higgs scalar with gravity and let the equations determine the dynamics of the scalar field and it's EOS. For this we use a a kinematic reconstruction from the cosmic statefinder parameter. Overall this serves us as a requirement of observational consistency of the late-time cosmology as well as a time varying Higgs VEV. \\

A requirement such as this is quite steep as well as unconventional. We rely on observational data from optical atomic clocks serving as the perfect tool to study deviations from the standard model of particle physics. In these observations, standard entities such as the Higgs VEV or the fine-structure constant are measured using light spectra from distant, extremely bright objects called quasars, which existed around $10$ billion years ago. If there is some variation of otherwise constant entities of the standard model with respect to cosmic time or redshift, the ticks of the atomic clocks will either speed up or slow down. We take into account the observational bound from these on $\frac{\Delta \nu}{\nu}$ or equivalently, on $- \frac{\Delta{\mu}}{\mu}$ using the data of cesium atomic clock \citep{huntemann} 
\begin{equation}
\frac{\Delta{\mu}}{\mu} = (-0.5 \pm 1.6)\times 10^{-16} \, \, year^{-1} .
 \label{017}
\end{equation}
The present Hubble constant is $H_{0} \simeq 7 \times 10^{-11}  \, \, year^{-1}$. Therefore one can write that roughly the variation has a scalae of the order $\frac{\Delta{\nu}}{\nu} \simeq 10^{-6} H_{0}$. Our primary goal is to constrain the varying Higgs VEV using the observational data written in Table. \ref{table1}. These are the weighted average of predominantly Hydrogen molecular spectra from different quasars \citep{king0, malec, weerd, wendt, dapra0, ubachs, kanekar}, listed in the form $\Delta \mu / \mu$ vs $z$.

\begin{table}
\caption{{\small Variation of $\Delta \mu / \mu$ from Hydrogen Molecular Spectra of Quasars for all $z > 2$ and other molecular spectra for $z < 1$.}}\label{table1}
\begin{tabular*}{\columnwidth}{@{\extracolsep{\fill}}lrrrrl@{}}
\hline
\multicolumn{1}{c}{Quasar} & \multicolumn{1}{c}{Redshift} & \multicolumn{1}{c}{$\Delta \mu / \mu \, \, [10^{-6}]$} \\
\hline
B0218+357 & 0.685 & $ -0.35  \pm 0.12 $ \\
PKS1830-211 & 0.89 & $ 0.08  \pm 0.47 $ \\
HE0027-1836 & 2.40 & $ -7.6  \pm 10.2 $ \\
Q0347-383 & 3.02 & $ 5.1  \pm 4.5 $ \\
Q0405-443 & 2.59 & $ 7.5  \pm 5.3 $ \\
Q0528-250 & 2.81 & $ -0.5  \pm 2.7 $ \\
B0642-5038 & 2.66 & $ 10.3  \pm 4.6 $ \\
J1237+064 & 2.69 & $ -5.4  \pm 7.2 $ \\
J1443+2724 & 4.22 & $ -9.5  \pm 7.5 $ \\
J2123-005 & 2.05 & $ 7.6  \pm 3.5 $ \\
\hline
\end{tabular*}
\end{table}

It is curious to note that Table \ref{table1} indeed suggests of a variation during late-time acceleration and the preceding deceleration, but no concrete claim can be made on a desired mathematical form of the evolution. Moreover, it remains to be seen if an analytical form chosen as a conjecture is consistent in comparison with the theory of gravity one choses as well as standard observations data e.g. the Luminosity-distance measurement data of Supernova or the Baryon Acoustic Oscillation data. We first claim that an interacting scalar field in the action of gravity is responsible for this variation. We work with three different forms of scalar extended theory of gravity in this regard. At the outset we simply assume that the aforementioned Higgs scalar field $\phi$ has a self-interaction of the form

\begin{equation}\label{potential}
V(\phi) = V_{0} + M(t) \phi^{2} + \frac{\lambda}{4} \phi^{4}.
\end{equation}

The scalar field is one dimensional in reduced Planck Mass unit and therefore the evolving coefficient $M(t)$ is of mass dimension two. $\lambda$ is dimensionless. From the $W-$boson mass ($M_{\small W}$) measurement bounds, we take that the Higgs VEV $\nu$ must be chosen such that
\begin{equation}
M_{\small W} \propto \nu \sim 246 GeV.
\end{equation}

From Eq. (\ref{potential}), the Higgs VEV $\nu$ can be calculated as
\begin{eqnarray}
&& \frac{\partial V}{\partial \phi} \bigg\rvert_{\nu} = 0,
\nu = \sqrt{\frac{-2M(t)}{\lambda}}.  \label{328}
\end{eqnarray}

We recall that the mass ratio $\frac{\Delta\mu}{\mu}$ evolves with redshift obeying 
\begin{equation}
\frac{\Delta \mu}{\mu}(z) = - \frac{91}{100} \frac{\nu(z) - \nu(z)\rvert_{z \sim 0}}{\nu(z)\rvert_{z \sim 0}},
\label{330}
\end{equation}

\begin{figure}
\begin{center}
\includegraphics[width=0.40\textwidth]{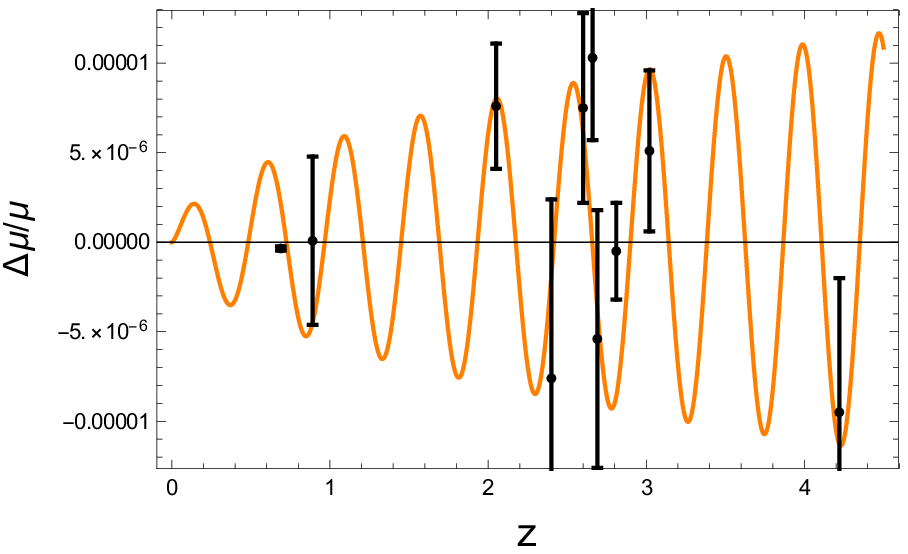}
\includegraphics[width=0.40\textwidth]{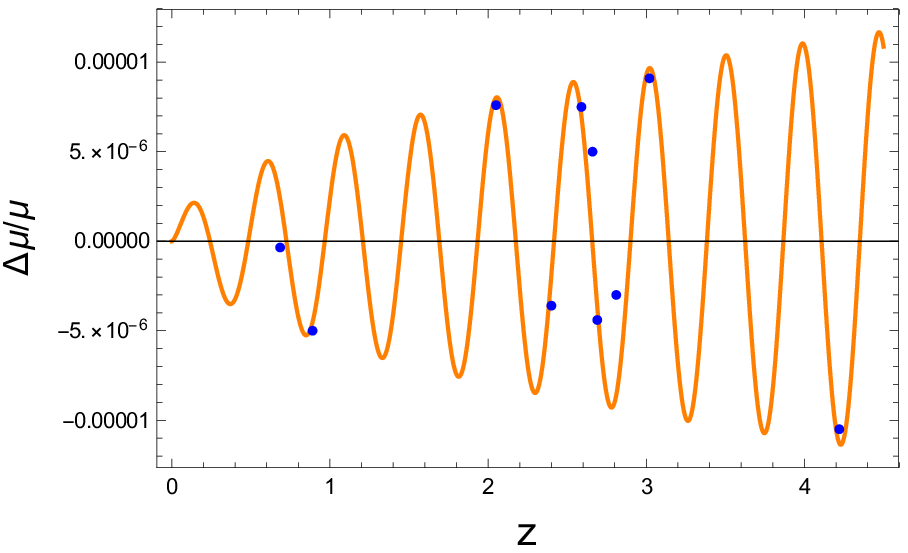}
\caption{Evolution of the conjectured $\Delta \mu / \mu $ as a function of redshift alongwith the fitted observational points from quasar absorption spectra, following Table \ref{table1}. The associated error bars are also shown in the graph below.}
\label{plot_1}
\end{center}
\end{figure}

and express $M(t)$ as a function of redshift for convenience to write
\begin{equation}
M(t) = M_{0} u(z).
\end{equation}  
The parameter $M_{0}$ has mass dimension two and the dimensionless $u(z)$ holds the key for the variation of $\mu$. We propose that $u(z)$ can be written in the following functional form

\begin{equation}\label{conjecture}
u(z) = - u_{1} + u_{2} (z + u_{3})^{u_{4}} sin (u_{5}z),
\end{equation}
where $u_{i}$-s are positive parameters. We calculate the mass ratio from this functional conjecture and fit with the Quasar absorption data (as in Table \ref{table1}) in Fig. \ref{plot_1}. Through some manual trials we estimate the required parameter values for a good fit. We also plot the exact data alongwith the associated error values in the bottom panel of the Figure. The estimated parameter values that give a good fit are
\begin{eqnarray}\nonumber
&& u_{1} \sim 0.2, \\&&\nonumber
u_{2} \sim 2.5 \times 10^{-6}, \\&&\nonumber
u_{3} \sim 0.001, \\&&\nonumber
u_{4} \sim 0.48, \\&&\nonumber
u_{5} \sim 13.
\end{eqnarray}

However, we do not claim this to be the only possible functional form that produces a mass ratio consistent with the Quasar spectra. This is only a particular example, atmost a toy model. The scope of this analysis is limited since observations from molecular absorption spectra aro not exactly accurate which can be understood by looking into the errors of each observational points. Irrespective of that, the implications of this very special case can not be denied since it can potentially serve as a new observational requirement for cosmological models, atleast in the low redshift regime. Moreover, with the rigorous advance of observational sector it may just be a matter of time before the variation of such fundamental constant entities are more profoundly established. Such examples are already in literature in the context of gravtational wave observations \citep{bagdon0}. 
 
\section{Late-time Acceleration and Constraints on the Scalar Dynamics}
In this section we focus on the cosmological aspects of this manuscript. We have a conjecture that the proton-to-electron mass ratio varies periodically with cosmic redshift in the low redshift regime. The scalar field and it's equation of state must have a non-trivial evolution in order to allow this variation as well as a consistent cosmology. Moreover, this will depend essentially how the scalar interacts with the theory of gravity. We discuss all of these using a cosmological reconstruction where the cosmic statefinder parameter is a primary tool. A reconstruction is a popular tool, primarily to avoid the complicacies of finding an exact solution for cosmological analysis. A reconstruction can come in different categories such as parametric and non-parametric approaches. It is also necessary to employ some form of statistical analysis for the comparison with a large set of data, for instance, a principle component analysis \citep{chen, ryan, critt, clarkson, ishida, amendola, hols, seikel, shafi}. Our method falls in the well-posed genre of kinematic reconstruction, as the statefinder is a purely kinematic parameter. Assuming that the statefinder parameter remains a constant during the late-time evolution of the universe, one can solve for the Hubble as a function of redshift $z$. This is followed by a comparison with observational data and subsequent estimation of the unknown functions and parameters of the theory. The reconstructed dynamical behavior of Hubble and the cosmological parameters are independent of the theory of gravity as these are simply reverse engineered from a kinematic quantity. We exploit this advantage and investigate if different scalar extensions of standard GR can support this reconstructed late-time cosmology. \\

Using the standard definitions of the Hubble and the Deceleration parameter, the statefinder parameter $s$ is defined as \citep{alam, sahni}
\begin{eqnarray}
r & = & \frac{\stackrel{\bf{...}}{a}}{aH^3} 
=\frac{\ddot H}{H^3}+3\frac{\dot H}{H^2}+1\label{eq:eq1.3} \\
s & = & \frac{r-1}{3\left(q-\frac{1}{2}\right)}, \label{eq:eq1.4}
\end{eqnarray}
A dot is cosmic time derivative and $a$ is the cosmological scale factor. Using a redefinition $1+z = \frac{1}{a} \equiv x$, the statefinder $s(x)$ can be written as a second order differential equation of Hubble,
\begin{equation}\label{mastereq1}
3 s \frac{H'}{H} x - \frac{9s}{2} = -2\frac{H'}{H}x + \frac{H'^2}{H^2}x^2 + \frac{H''}{H}x^2.
\end{equation}

We rely on a constant statefinder parameter to describe the deceleration to acceleration transition of the universe, the viability of which has been discussed quite recently \citep{chakrabarti}. One must keep in mind here that for a standard $\Lambda$CDM cosmology, the statefinder is zero. Thus, a non-zero constant statefinder essentially is a measure of allowed departure from $\Lambda$CDM. A better form of this reconstruction would be to treat the statefinder as a function rather than a constant, however, we keep such a mathematical challenge for a future work. Writing $s = \delta - \frac{2}{3}$, Eq. (\ref{mastereq1}) can be solved to write 

\begin{eqnarray}\nonumber\label{hubblex}
&& H(x) = C_{2}x^{\frac{1}{4}\left[1+3\delta- \left\lbrace 1 + 6\delta + 9\delta^2 + 36\left(\delta - \frac{2}{3}\right)\right\rbrace^{1/2}\right]} \\&&
\left[C_{1} + x^{\left\lbrace 1 + 6\delta + 9\delta^2 + 36\left(\delta - \frac{2}{3}\right)\right\rbrace^{1/2}}\right]^{1/2}.
\end{eqnarray}

The two integration constants $C_{1}$ and $C_{2}$ serve different roles. While $C_{2}$ is connected to the present Hubble value, $C_{1}$ must be estimated from observational data. From Eq. (\ref{hubblex}) we ensure that the evolution is real and demand that $\left\lbrace 1 + 6\delta + 9\delta^2 + 36\left(\delta - \frac{2}{3}\right)\right\rbrace$ must be positive. This implies a constraint on $\delta$ as
\begin{equation}
\delta^2 + \frac{14}{3}\delta - \frac{23}{9} > 0.
\end{equation} 

A model described by Eq. (\ref{hubblex}) can successfully produce the signature flip of deceleration $q$. We confirm this by comparing with observational data and study the confidence contours and estimate the best fit values of the model parameters. We use data sets from (a) Joint Light-Curve Analysis data \citep{jla} of luminosity distance measurement ($SDSS-II$ and $SNLS$ collaborations), the Hubble parameter measurements (OHD) \citep{ohd1, ohd2, ohd3, ohd4, ohd5, ohd6, delu, planck} and the Baryon Acoustic Oscillation (BAO) data \citep{6dF, bossanderson}. Changing the variable using $1 + z = \frac{1}{a} \equiv x$, we rewrite the Hubble as a function of redshift
\begin{eqnarray}\nonumber\label{hubblez}
&& H(z) = \\&&\nonumber
\frac{H_{0}}{(1+C_{1})^{1/2}}(1+z)^{\frac{1}{4}\left[1+3\delta- \left\lbrace 1 + 6\delta + 9\delta^2 + 36\left(\delta - \frac{2}{3}\right)\right\rbrace^{1/2}\right]} \\&&
\left[C_{1} + (1+z)^{\left\lbrace 1 + 6\delta + 9\delta^2 + 36\left(\delta - \frac{2}{3}\right)\right\rbrace^{1/2}}\right]^{1/2}.
\end{eqnarray}

\begin{figure}
\begin{center}
\includegraphics[angle=0, width=0.52\textwidth]{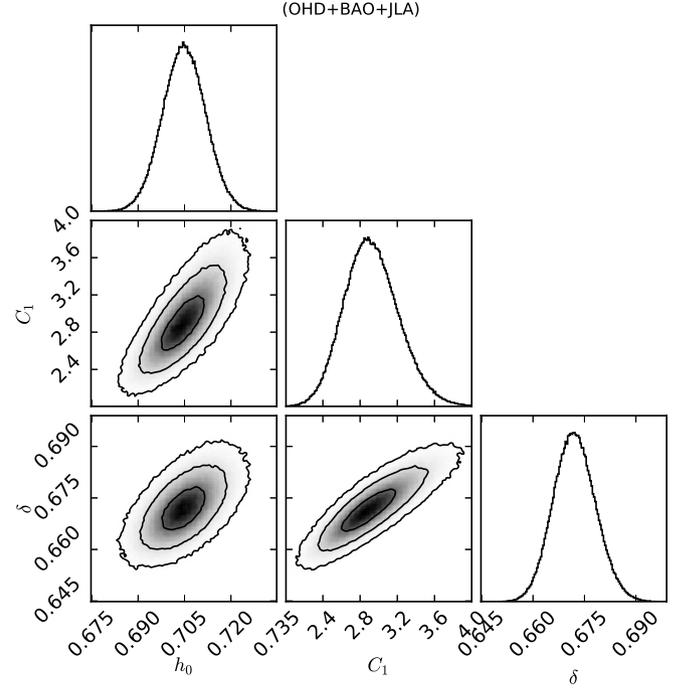}
\caption{Estimation of three parameters, confidence contours and marginalized likelihood function using OHD+JLA+BAO data.}
\label{Modelcontour}
\end{center}
\end{figure}

\begin{table}
\caption{{\small The model parameter estimation alongwith 1$\sigma$ uncertainty.}}\label{resulttable}
\begin{tabular*}{\columnwidth}{@{\extracolsep{\fill}}lrrrrl@{}}
\hline
 & \multicolumn{1}{c}{$h_0$} & \multicolumn{1}{c}{$C_{1}$} & \multicolumn{1}{c}{$\delta$} \\
\hline
$OHD+JLA+BAO$ 	  & $0.705^{+0.007}_{-0.007}$ &$2.913^{+0.303}_{-0.274}$ & $0.672^{+0.006}_{-0.006}$ &\\
\hline
\end{tabular*}
\end{table}

\begin{figure}
\begin{center}
\includegraphics[angle=0, width=0.40\textwidth]{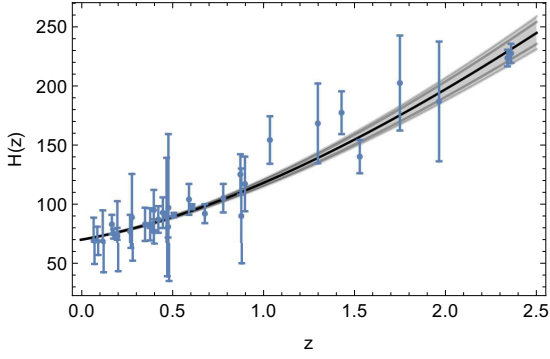}
\caption{$H(z)$ vs $z$ : The best fit (thick black line) evolution, associated 2$\sigma$ and 3$\sigma$ regions (the gray zones).}
\label{Hz_data}
\end{center}
\end{figure}

We compare our theoretical ansatz with observational data using statistical analysis. Primarily we estimate three entities, (a) the dimensionless Hubble $h_{0} = H_{0}/100 km\mbox{Mpc}^{-1} \mbox{sec}^{-1}$, (b) the evolution and the present value of deceleration parameter and (c) the redshift where the deceleration parameter switches sign from positive to negative. For the comparison and subsequent parameter estimation we use a Markov Chain Monte Carlo simulation in python (the EMCEE \citep{foreman}). The code allows us to estimate the parameter values for which one gets the best cosmological behavior, alongwith error estimations (See Table \ref{resulttable}). The Hubble parameter at the present epoch is estimated to be $70.5 km\mbox{Mpc}^{-1} \mbox{sec}^{-1}$, which goes very well with recent observations \citep{riess2018}. In Fig. \ref{Hz_data} the evolution of $H(z)$ is shown alongwith 2$\sigma$ and 3$\sigma$ error estimation regions. Using the best fit value of $\delta$, we estimate that the statefinder parameter must be in the range $0.011 \geq s \geq -0.001$. This essntially points out that a consistent statefinder diagnostic demands a late-time cosmology very close to $\Lambda$CDM.  \\

\begin{figure}
\begin{center}
\includegraphics[width=0.40\textwidth]{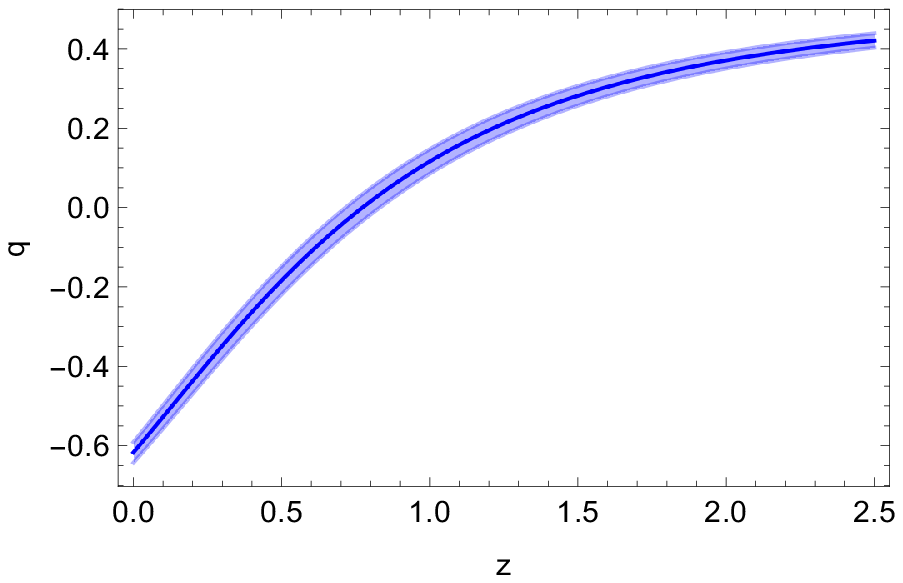}
\includegraphics[width=0.40\textwidth]{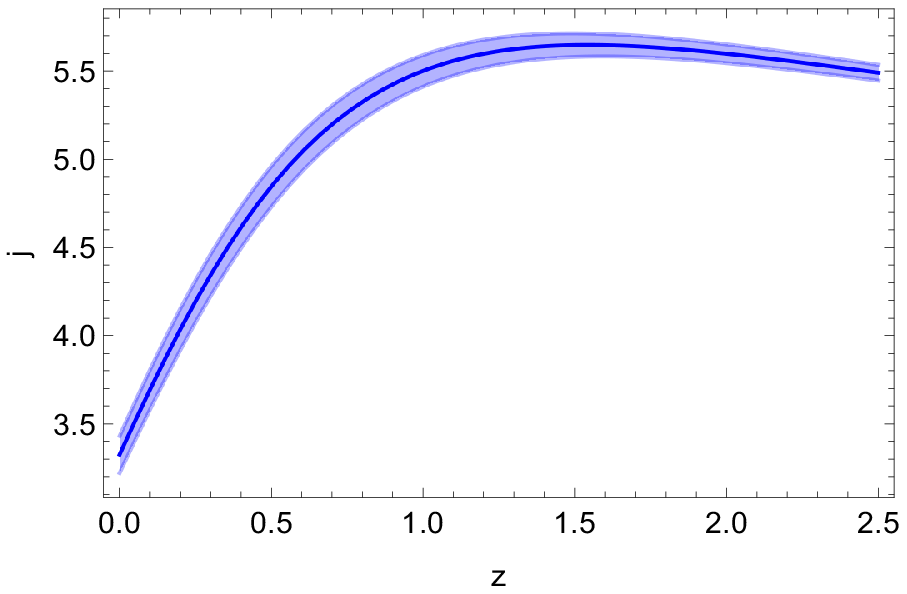}
\caption{Top graph : Deceleration $q(z)$ vs $z$. Bottom graph : Jerk $j(z)$ vs $z$. The best fit plot is in bold blue and the 3$\sigma$ region is in faded blue.}
\label{kinematic_parameters}
\end{center}
\end{figure}

The signature flip of the deceleration $q(z)$ is shown in the top panel of Fig. \ref{kinematic_parameters}. The bold blue curve is for best fit values and the faded blue area is the 3$\sigma$ confidence region. At $z \sim 0$ the deceleration parameter $q(z)$ is $\sim -0.62$. The transition of $q(z)$ from positive into negative is realized around a redshift of transition $z_{t} < 1$. Both of these results are observationally very well-consistent \citep{riess, farooq}. For a standard $\Lambda$CDM, the jerk parameter is $1$. However, in the present case there is a clear deviation which is evident from the bottom graph of Fig. \ref{kinematic_parameters}. Therefore, we comment that the requirement of observational consistency can support non-trivial deviation from standard GR, but the deviation is only realized through higher order kinematic variables.    \\

We recall that the reconstruction is based on just a kinematical quantity. Therefore the reconstructed quantities do not depend on the theory of gravity until one chooses a matter distribution and employs the field equations. For instance, in the context of GR, the equation of state (EOS) as a function of $z$ is

\begin{figure}
\begin{center}
\includegraphics[width=0.40\textwidth]{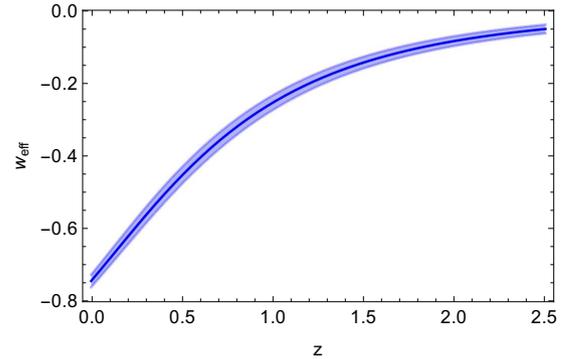}
\caption{Equation of state vs $z$ : bold blue curve gives the best fit and the faded blue curve gives the 3$\sigma$ region.}
\label{eosfig}
\end{center}
\end{figure}

\begin{eqnarray}\nonumber
&&w_{eff}=\frac{p_{tot}}{\rho_{tot}},\\&&\nonumber
\frac{\rho_{tot}}{\rho_{c0}} = \frac{H^2(z)}{H^2_0}, \\&&\nonumber
\frac{p_{tot}}{\rho_{c0}}=-\frac{H^2(z)}{H^2_0}+\frac{2}{3}\frac{(1+z)H(z)H'(z)}{H^2_0}, \\&&\nonumber
\rho_{c0}=3H^2_0/8\pi G.
\end{eqnarray}

In Fig. \ref{eosfig} we plot $w_{eff}$ across two epochs. For low redshift the effective EOS $w_{eff}$ is negative, which is a necessary condition to drive an accelerated expansion ($\rho_{c0}$ being critical density). With $z$, $w_{eff}$ shows an increasing behavior until it reaches around zero. This essentially means that the present dark energy dominated acceleration is preceded by a dust dominated decelerated epoch. Using the Hubble expression we can also determine if the present universe is in thermodynamic equilibrium or not. If the cosmological system follows blackhole thermodynamics \citep{gibbons, jacobson, bakrey}, for a thermodynamic equilibrium the total entropy $S = S_f + S_h$ (sum of fluid entropy $S_f$ and  boundary entropy on Hubble horizon $S_h$) must follow 
\begin{eqnarray}
&& \frac{dS}{dn}\geq 0, \\&&
\frac{d^2S}{dn^2} < 0, \\&&
n=\ln{a}. \label{therm}
\end{eqnarray}
The derivatives of the entropy are related to the Hubble and it's derivatives through \citep{jamil, supriyapan}

\begin{eqnarray}\label{Entropy_Psi}
&& S_{,n} \propto \frac{(H_{,n})^2}{H^4}, \\&&
S_{,nn} = 2S_{,n}\left(\frac{H_{,nn}}{H_{,n}}-\frac{2H_{,n}}{H}\right) = 2S_{,n}\Psi.
\end{eqnarray}

\begin{figure}
\begin{center}
\includegraphics[width=0.40\textwidth]{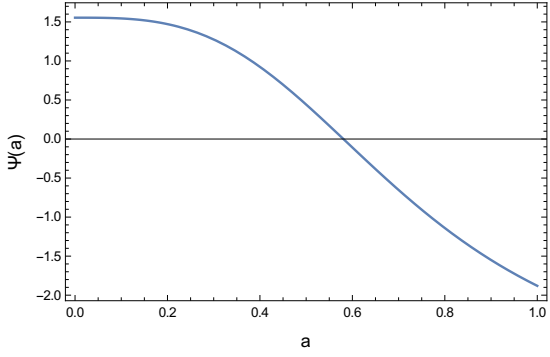}
\caption{Thermodynamic equilibrium of the system : $\Psi$ vs $a$ for the best fit parameter values.}
\label{psiplot}
\end{center}
\end{figure}

A thermodynamic equilibrium is determined by the signature of $\Psi$. $\Psi$ as a function of $a$ clarifies the scenario, given in Fig. \ref{psiplot}. There is no discontinuity in the transition of $\Psi$ across different values and a late time acceleration is associated with thermodynamic equilibrium ($\Psi < 0$).

Thus we have essentially found a way to standardize a viable late-time cosmology with close resemblance with $\Lambda$CDM without the need to specify any theory of gravity at the outset. Cosmological viability of this kinematic approach can be addressed at a greater length, by discussing for instance the growth of matter overdensity. For such details, please refer to the work of \cite{chakrabarti}. For the present manuscript the reconstruction is not the main perspective. We do have an analytical form of Hubble and this is enough for us to look into three examples of scalar extended theories of gravity. Our aim is to discuss the evolution of the scalar dark energy as well as the effective equation of state of the system driving the acceleration of the universe. 

\subsection{Minimally Coupled Scalar Cosmology}
The first case we consider is a theory of gravity where a spatially homogeneous scalar field is minimally coupled in the lagrangian. This makes the action
\begin{equation}\label{action1}
\textit{A} = \int{\sqrt{-g} d^4x [R + \frac{1}{2}\partial^{\mu}\phi\partial_{\nu}\phi - V(\phi) + L_{m}]}.
\end{equation}
$L_{m}$ signifies that there is a fluid energy momentum distribution alongwith the scalar field. In natural units, the field equations are
\begin{equation} \label{fe1minimal}
3\Big(\frac{\dot{a}}{a}\Big)^{2} = \rho_{m} + \rho_{\phi} = \rho_{m} + \frac{\dot{\phi}^{2}}{2} + V\left( \phi \right),
\end{equation} 
 
\begin{equation} \label{fe2minimal}
-2\frac{\ddot{a}}{a}-\Big(\frac{\dot{a}}{a}\Big)^{2} = p_{m} + p_{\phi} = p_{m} + \frac{\dot{\phi}^{2}}{2}-V\left( \phi \right),
\end{equation}

alongwith the scalar evolution equation
\begin{equation} \label{phiminimal}
\ddot{\phi} + 3\frac{\dot{a}}{a}\dot{\phi} + \frac{dV(\phi)}{d\phi} = 0.  
\end{equation}     

As discussed in Section $2$ we take the scalar interaction in it's non-trivial Higgs potential form as in Eq. (\ref{potential}) 
\begin{equation}
V(\phi) = V_{0} + M(t) \phi^{2} + \frac{\lambda}{4} \phi^{4}.
\end{equation}

With this, the scalar evolution in Eq. (\ref{phiminimal}) becomes
\begin{equation}\label{minKG}
\ddot{\phi} + 3\frac{\dot{a}}{a}\dot{\phi} + \lambda\phi^3 + 2\phi M(z) + \phi^{2}\frac{\frac{\partial M}{\partial z}}{\frac{\partial \phi}{\partial z}}= 0.
\end{equation}

We take $M = M_{0}u(z)$, $\phi = \phi_{0} \psi(z)$ ($\phi_{0}$ is of dimension $M_{p}$) and transform Eq. (\ref{minKG}) into a function of scale factor and redshift using the kinematic reconstruction of Hubble as in Eq. (\ref{hubblex}) and write

\begin{figure}
\begin{center}
\includegraphics[width=0.40\textwidth]{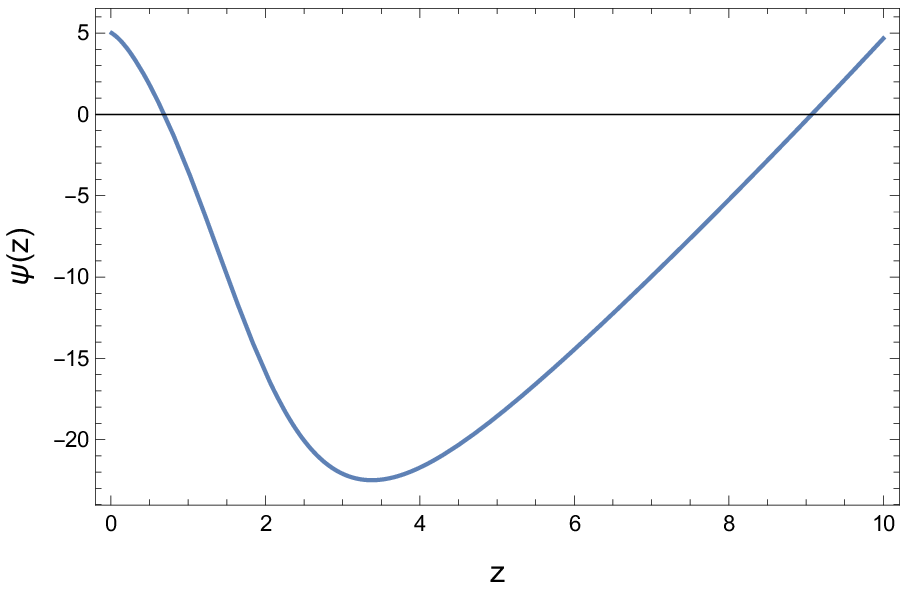}
\includegraphics[width=0.40\textwidth]{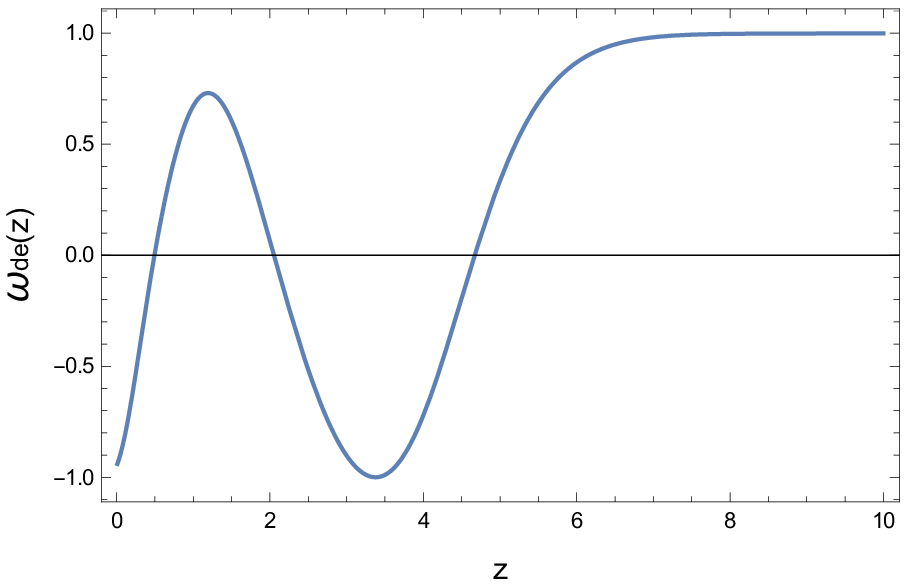}
\includegraphics[width=0.40\textwidth]{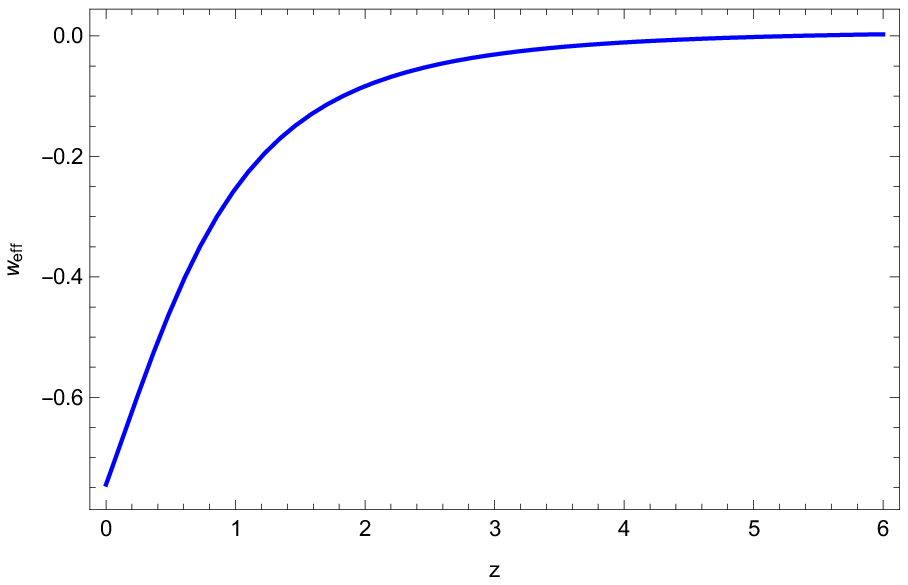}
\caption{Minimially Interacting Higgs Scalar : Evolution of the scalar field, $\omega_{de}$ and $\omega_{eff}$ as a function of redshift. The initial condition for numerical solution is $\phi_{z=0} > 0$ and $\frac{d\phi}{dz}_{z=0} < 0$.}
\label{plot_minimallycoupled2}
\end{center}
\end{figure}

\begin{eqnarray}\label{minKGredshift} \nonumber
&& \psi^{\circ\circ} = \psi^{\circ} a + \psi^{\circ} a^{\frac{17}{20}} \Big(3 + a^{-3.06}\Big)^{-\frac{1}{2}} \Bigg[(3 + a^{-3.06})^{\frac{1}{2}} \\&&\nonumber
-\frac{3}{2}\Big(3 + a^{-3.06}\Big)^{-\frac{1}{2}}a^{-3.06}\Bigg] -\frac{M_{0}\psi^{2}u^{\circ}}{C_{2}^{2}\psi^{\circ}}a^{\frac{17}{10}}\\&&\nonumber
\Big(3 + a^{-3.06}\Big)^{-1} - \frac{2M_{0}}{C_{2}^{2}}a^{\frac{17}{10}}\Big(3 + a^{-3.06}\Big)^{-1}u -\\&&
\frac{\lambda \phi_{0}^{2}\psi^{3}}{C_{2}^{2}}a^{\frac{17}{10}}(3 + a^{-3.06})^{-1}.
\end{eqnarray}
$\psi^{\circ}$ is derivative of $\psi$ with respect to $z$. \\
\begin{figure}
\begin{center}
\includegraphics[width=0.40\textwidth]{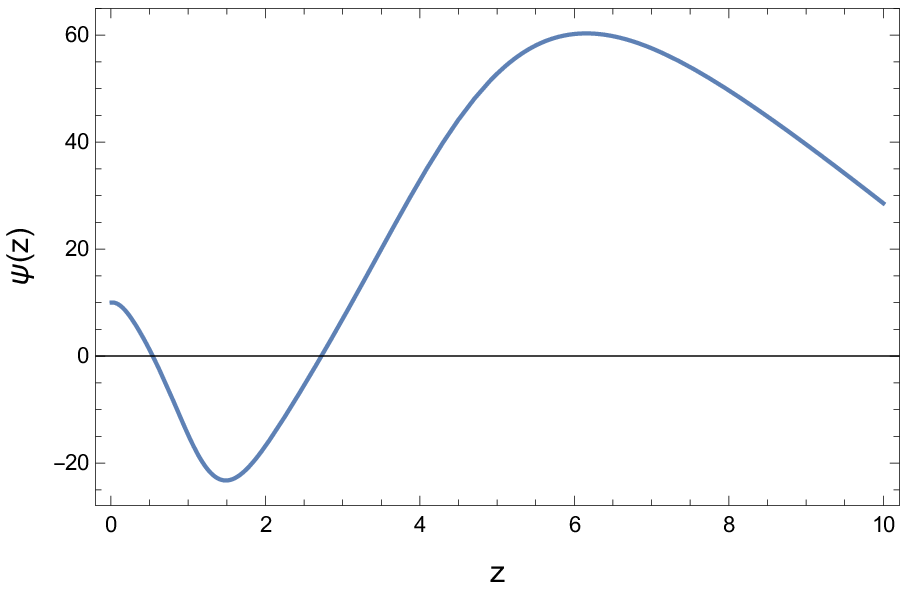}
\includegraphics[width=0.40\textwidth]{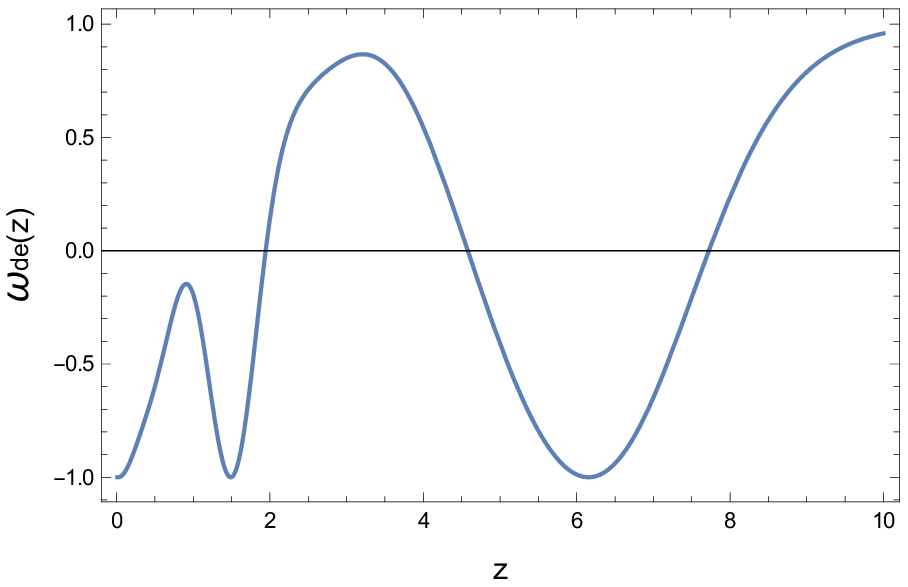}
\includegraphics[width=0.40\textwidth]{weff_z.eps}
\caption{Minimially Interacting Higgs Scalar : Evolution of the scalar field, $\omega_{de}$ and $\omega_{eff}$ as a function of redshift. The initial condition for numerical solution is $\phi_{z=0} > 0$ and $\frac{d\phi}{dz}_{z=0} > 0$.}
\label{plot_minimallycoupled3}
\end{center}
\end{figure}

Eq. (\ref{minKGredshift}) governs the scalar evolution in different epochs. The first step of demonstration is the numerical solution of this equation for two sets of initial conditions. The second step is to solve the two independent field equations Eq. (\ref{fe1minimal}) and Eq. (\ref{fe2minimal}) for two unknowns, the perfect fluid density and pressure. It is a bit lengthy but straightforward to write these equations as a function of redshift using Eq. (\ref{hubblex}) and we choose not to repeat a similar procedure again and again. In all of our constructions here in this manuscript, we expect the scalar field to behave as the dark energy and the accompanying fluid description to behave as a dark matter. Therefore, numerical solution of the complete system must produce an evolving equation of state of the scalar dark energy as well as an evolving effective equation of state for the entire system. Figs. \ref{plot_minimallycoupled2} and \ref{plot_minimallycoupled3} contain the plots of these quantities as a function of redshift. The two figures are for two different sets of initial condition, $\frac{d\phi}{dz}_{i} < 0$ and $\frac{d\phi}{dz}_{i} > 0$ where the initial scalar field value is chosen positive for both of the cases. While a negative initial value for the scalar field is perfectly reasonable to consider, it does not produce any qualitative difference in the numerical solutions, other than some scaling. We can see that for the Higgs VEV to take an oscillatory form, the scalar field as well as the equation of state of the scalar dark energy must show a hint of oscillatory behavior, atleast in the low redshift regime. This is more prominent, in Fig. \ref{plot_minimallycoupled3} with the initial condition $\frac{d\phi}{dz}_{i} > 0$. However, the interesting outcome of this construction is that this periodicity hardly leaves any signature on the effective equation state of the non-interacting scalar-fluid system. The evolution of the effective equation state is shown in the bottom panel of each of the figures and it clearly suggests a standard dark energy dominated universe around $z \sim 0$ and a zero approaching $\omega_{eff}$ indicating a dust dominated universe prior to the present acceleration. Thus we can take a hint from here that a non-trivial oscillatory behavior of scalar dark energy component during cosmic evolution can be suppressed due to the presence of a perfect fluid in the system. 

\subsection{Non-Minimally Coupled Scalar Cosmology}
However, we expect that inclusion of scalar interactions with geometry can have a modified impact on the signature of varying Higgs VEV. We discuss this in the context of non-minimally coupled scalar theories of cosmology which are written by allowing a direct scalar interaction with the lagrangian \citep{chernikov, birrell}. These theories have added motivation of being able to renormalize scalar field theories with quantum corrections \citep{dono, ford, callan}, also in the context of superstring theory \citep{maeda}. In a cosmological backdrop, non-minimally coupled scalars are expected to play a bigger role compared to their `increasingly-redundant' minimally coupled counterparts. The special case of a quadratic scalar interaction is particularly favoured in this regard \citep{martin, luo, szyd, atkins} and we consider a similar setup. We first write the action as

\begin{equation}
S =\int d^4x\sqrt{-g}\left[W(\phi)R - \frac{1}{2}g^{\mu\nu}\phi_{,\mu}\phi_{,\nu} + V(\phi) + L_{m}\right].
\label{nonminaction}
\end{equation}

Usual variation with the metric and the scalar field produces the cosmological field equations
\begin{equation}
6 W H^{2} + 6 W'H\dot{\phi} = \rho_{m} + \frac{1}{2}\dot{\phi}^2 + V(\phi),
\label{nonminfrw1}
\end{equation}

\begin{equation}
4W\dot{H} + 6WH^2 + 4W'H\dot{\phi} + 2W''\dot{\phi}^2 + 2W'\ddot{\phi} = -p_{m} -\frac{\dot{\phi}^2}{2} + V(\phi).
\label{nonminimalfrw2}
\end{equation}

and the scalar equation
\begin{equation}
\ddot{\phi} + 3H\dot{\phi} - 6W'\left[\dot{H} + H^{2}\right] + \frac{dV}{d\phi} = 0.
\label{nonminKGini}
\end{equation}

The scalar-geometry non-minimal interaction is taken to be quadratic in nature and written as
\begin{equation}
W(\phi) = \frac{1}{2}(1 + U_{0}\phi^2).
\end{equation}

As in the minimally coupled scalar evolution, once again we take the self-interaction of the scalar to be Higgs type as in Eq. (\ref{potential}) and write the scalar evolution equation as 

\begin{equation}\label{nonminKG}
\ddot{\phi} + 3H\dot{\phi} - 6 U_{0} \phi \left[\dot{H} + H^{2}\right] + \lambda\phi^3 + 2\phi M(z) + \phi^{2}\frac{\frac{\partial M}{\partial z}}{\frac{\partial \phi}{\partial z}} = 0.
\end{equation}

We put $M = M_{0}u(z)$, $\phi = \phi_{0} \psi(z)$ ($\phi_{0}$ is of dimension $M_{p}$) and use Eq. (\ref{hubblex}) to write the above Eq. (\ref{nonminKG}) as

\begin{figure}
\begin{center}
\includegraphics[width=0.40\textwidth]{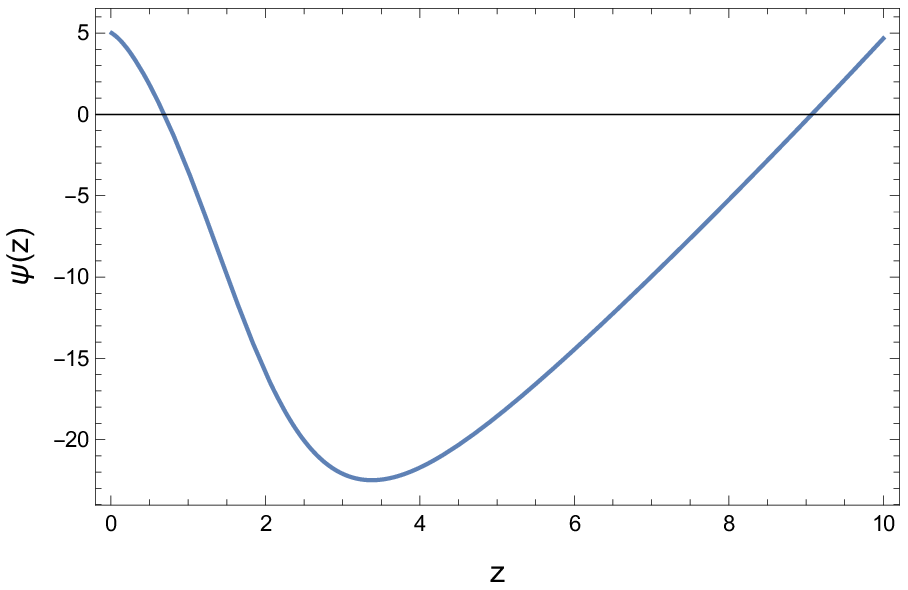}
\includegraphics[width=0.40\textwidth]{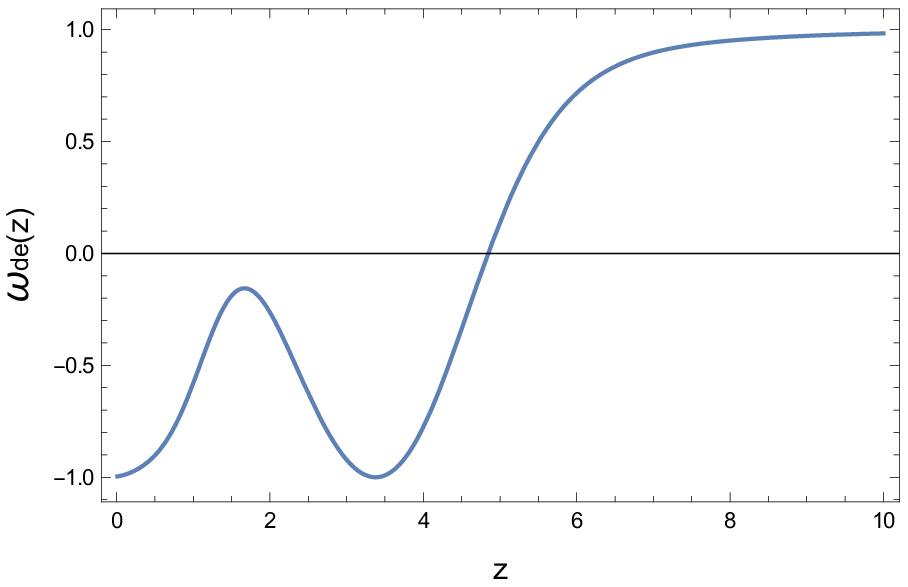}
\includegraphics[width=0.40\textwidth]{weff_z.eps}
\caption{Nonminimially Interacting Higgs Scalar : Evolution of the scalar field, $\omega_{de}$ and $\omega_{eff}$ as a function of redshift. The initial condition for numerical solution is $\frac{d\phi}{dz}_{z=0} < 0$.}
\label{plot_nonminimallycoupled2}
\end{center}
\end{figure}

\begin{figure}
\begin{center}
\includegraphics[width=0.40\textwidth]{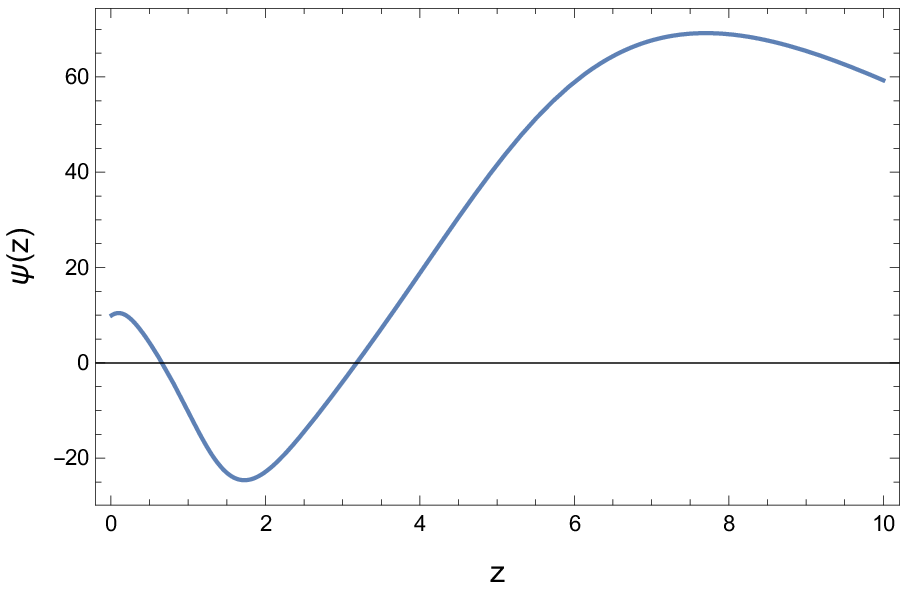}
\includegraphics[width=0.40\textwidth]{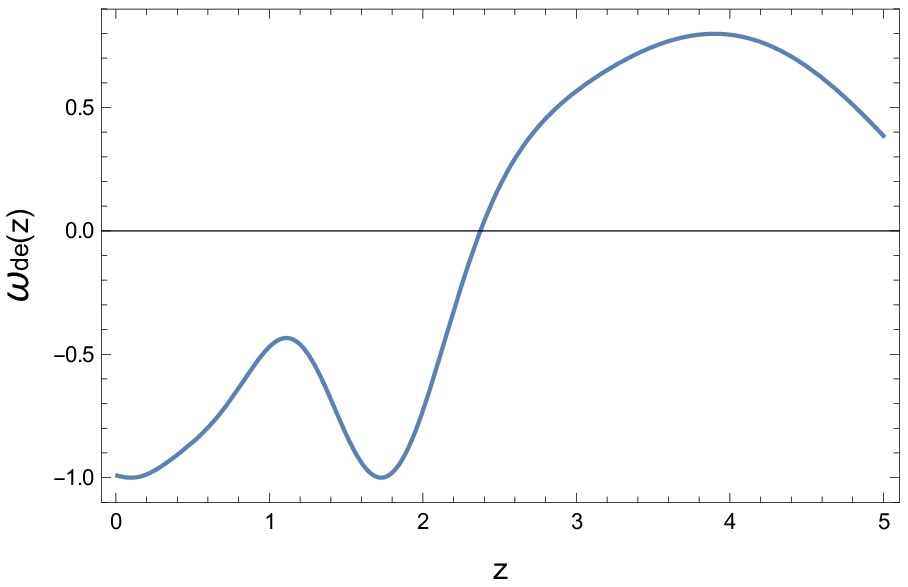}
\includegraphics[width=0.40\textwidth]{weff_z.eps}
\caption{Nonminimially Interacting Higgs Scalar : Evolution of the scalar field, $\omega_{de}$ and $\omega_{eff}$ as a function of redshift. The initial condition for numerical solution is $\frac{d\phi}{dz}_{z=0} > 0$.}
\label{plot_nonminimallycoupled3}
\end{center}
\end{figure}

\begin{eqnarray}\label{nonminKGredshift} \nonumber
&& \psi^{\circ\circ} = \psi^{\circ} a + \psi^{\circ} a^{\frac{17}{20}} \Big(3 + a^{-3.06}\Big)^{-\frac{1}{2}} \Bigg[\Big(3 + a^{-3.06} \Big)^{\frac{1}{2}} \\&&\nonumber
-\frac{3}{2}\Big(3 + a^{-3.06}\Big)^{-\frac{1}{2}}a^{-3.06}\Bigg] + 6U_{0}\psi \Bigg[\Big(3 + a^{-3.06}\Big)\\&&\nonumber
\Bigg\lbrace\Big(3 + a^{-3.06}\Big)^{\frac{1}{2}} - \frac{3}{2}a^{-3.06}\Big(3 + a^{-3.06}\Big)^{-\frac{1}{2}}\Bigg\rbrace + a^{\frac{3}{10}}\\&&\nonumber
\Big(3 + a^{-3.06}\Big) \Bigg]a^{\frac{17}{10}}\Big(3 + a^{-3.06}\Big)^{-1} - \frac{M_{0}\psi^{2}u^{\circ}}{C_{2}^{2}\psi^{\circ}}a^{\frac{17}{10}}\\&&\nonumber
\Big(3 + a^{-3.06}\Big)^{-1} - \frac{2M_{0}u}{C_{2}^{2}}a^{\frac{17}{10}}\Big(3 + a^{-3.06}\Big)^{-1} -\frac{\lambda \phi_{0}^{2}\psi^{3}}{C_{2}^{2}}\\&&
a^{\frac{17}{10}}(3 + a^{-3.06})^{-1},
\end{eqnarray}
where instead of cosmic time, redshift and scale factor are the main variables. Similar to the previous subsection, Eq. (\ref{nonminKGredshift}) is the equation to solve for a clear picture of the scalar evolution. The numerical solution is done for two sets of initial conditions, $\frac{d\phi}{dz}_{z=0} < 0$ and $\frac{d\phi}{dz}_{z=0} > 0$. The two independent field equations Eq. (\ref{nonminfrw1}) and Eq. (\ref{nonminimalfrw2}) serve as constraint equations, with two unknowns to be determined, the perfect fluid density and pressure. For the sake of brevity we do not go into detailed steps and straightaway present the results, which are, the equation of state of the scalar dark energy component, an evolving effective equation of state of the entire system and the scalar evolution. Figs. \ref{plot_nonminimallycoupled2} and \ref{plot_nonminimallycoupled3} show these quantities as a function of redshift for the two different sets of initial condition where the initial scalar value is chosen positive for both of the cases. For the numerical calculation we have chosen $U_{0} = \frac{1}{6}$. The nature of the theoretical construct is found to be almost similar to a minimally coupled case,  with some important qualitative differences. For the Higgs VEV to take an oscillatory form, the scalar field as well as the equation of state of the scalar dark energy must take an oscillatory form in the low redshift regime. In this case as well, the oscillatory behavior of the scalar hardly leaves any signature on the effective equation state of the scalar-fluid system. The effective equation state is shown in the bottom panel of each of the figures and it clearly suggests a standard dark energy dominated universe around $z \sim 0$. With redshift, $\omega_{eff}$ approaches zero indicating a dust dominated universe prior to the present acceleration. Thus we once again show that, the non-trivial oscillatory behavior of the scalar dark energy required to support a varying Higgs VEV is supposed to remain well suppressed due to the presence of an additional perfect fluid description. \\

\begin{figure}
\begin{center}
\includegraphics[width=0.40\textwidth]{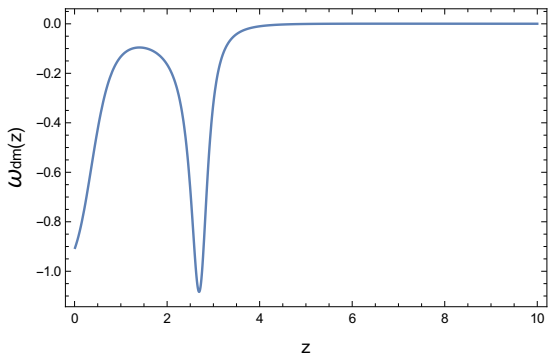}
\caption{Evolution of the perfect fluid ($\omega_{dm}$) EOS as a function of redshift.}
\label{fluideos}
\end{center}
\end{figure}

In this regard, it is quite important to atleast mention the nature of the accompanying perfect fluid in these models. It plays a crucial role in supporting the nontrivial sinusoidal evolution of the Higgs VEV across cosmological epochs. The generic nature is shown in Fig. \ref{fluideos} and shows a more dominant sign of periodicity compared to the scalar EOS, atleast around the redshift of transition from deceleration into acceleration. In a region of higher redshift, the $\omega_{fluid}$ (or the $\omega_{dm}$) approaches zero, pointing towards a dust-like behavior. Therefore it is more likely to find the signature of a varying proton-to-electron mass ratio or any fundamental constant entities, in a better description of the cosmological dark matter.

\subsection{Interacting Brans-Dicke-Higgs Theory}
The standard scalar field models discussed in the last two subsections are straightforward examples to employ a scalar field to drive the late-time cosmology. The scalar fields in these setups have self-interaction and in one case, an interaction with the curvature scalar in the lagrangian. However, these fields do not have any interaction with matter distribution as in the accompanying perfect fluid. In a sense these are cosmological systems with a non-interacting mix of scalar dark energy and a fluid tipped to be the dark matter. This serves cosmological requirements nicely. But in light of recent observations and to address the unresolved loose ends of cosmology, a more appropriate route requires more generalizations. The first step is to propose that the scalar Dark Energy component has a varying strength, i.e., it can remain latent during deceleration and comes to dominate only in late-times. Even in simple minimally coupled Quintessence scalar cosmology, the scalar potential dominates over a dormant kinetic term during the late time acceleration \citep{copeland}. The second is to replace simple ad-hawk scalar fields with field theory inspired interacting scalar fields, as in a more complete approach of Scalar-Tensor theories. The Brans-Dicke (BD) \citep{bransdicke} theory has served the GR community for about half a century as the prototype of Scalar-Tensor theories. The theory has been generalized/extended in more than a few ways over the years and these extensions have their fair share of successes and failures from a cosmological purview \citep{bergmann, nordtvedt, wagoner, barker, santos, holden, bertolami, nbsudipta, faraoni, mathia, la, nbpavon, mota, clifton}. Third and most crucially, many of these arguments are in favor of an interacting Dark Energy-Dark Matter scenario which is receiving increasingly attention as a better driver of the cosmological evolution as well as a resolution of the so-called coincidence problem.  \\

In this subsection, we explore the possibility of supporting a varying Higgs VEV in a two-scalar extended BD theory. The setup, apart from the standard BD scalar, incorporates a Higgs scalar field $\phi$ with self-interaction as in Eq. (\ref{potential}). The other interaction terms included in the action are the standard interaction of the BD field $\psi$ with curvature, derivative and non-derivative interaction of $\phi$ with curvature and an interaction between $\phi$ and $\psi$. Inspite of being complicated, such an action highlights the interplay between particle physics and GR and produces interesting physics, such as a Higgs driven inflation \citep{bezrukov, germani, masina, tsuji, alexander} and observationally consistent Higgs cosmology. These models are also compatible with LHC observations \citep{atkins, onofrio}. A similar setup of scalar extended GR can indeed support a Higgs scalar driven power-law cosmology, with a mildy varying VEV \citep{sola0}. Although power-law solutions give a constant deceleration parameter and are ruled out in view of observations, these can easily be thought of as asymptotic states of an unified cosmic evolution with evolving deceleration \citep{amendola00}. It is also possible to extend the theory and write an extended Brans-Dicke-Higgs setup to describe a consistent late-time cosmology \citep{chakrabarti}. Further generalizations have also led to the foundation of `running vacuum $\Lambda$CDM' models \citep{sola2, sola3, sola4}. We take it at the outset that there is a varying Higgs VEV in our cosmological setup and reconstruct the profiles of the BD scalar, Higgs scalar, their interactions and the EOS parameters as a function of redshift. Cosmological consistency is achieved by the statefinder diagnostic discussed before, using primarily Eq. (\ref{hubblez}). The action of the theory \citep{chakrabarti} is

\begin{eqnarray}\nonumber
&& S = \int d^{4}x\sqrt{-g}\Bigg[\frac{1}{2}R\psi - \frac{\omega}{2\psi}g^{\mu\nu}\partial_{\nu}\psi\partial_{\mu}\psi + U(\phi , \psi) \\&& 
+ \xi R\phi^{2} - \frac{1}{2} g^{\mu\nu}\partial_{\mu}\phi\,\partial_{\nu}\phi + \frac{1}{\phi^{2}}S_{\mu\nu}\partial^{\mu}\phi\partial^{\nu}\phi - V(\phi)\Bigg].  \label{eq:SMBDaction}
\end{eqnarray}

We neither take any interaction of the BD field nor any fluid energy momentum tensor. We expect the two scalars to serve the prospect of describing an interacting dark metter-dark energy scenario. The similarity in the order of magnitude of the two dark components at the present era seems to support a correlation and is additionally inspired from theories of unification \citep{kamen, bento0, bilic, farrar, bertolamiinter}. We also assume that the geometric scalar has an overall mild variation with time for the Weak Equivalence Principle to hold \citep{damour}. The non-trivial derivative interaction of $\phi$ with the action is inspired from certain models of inflation, scalar formulation of varying $G$ theories and also scalar quantum electrodynamics \citep{amendola0, capolamb}. 

\begin{eqnarray}\label{eq:SmunuTensor}
&& S_{\mu\nu}\equiv\varsigma R_{\mu\nu}-\frac{\theta}{2}g_{\mu\nu}R, \\&&
\theta \neq \varsigma.
\end{eqnarray}

We rescale the gravitational interaction by writing a rescaled BD field $\psi \sim 8\pi\psi$

\begin{equation}\label{eq:Geff}
G_{eff}(t) = \frac{1}{8\pi\psi(t)}.
\end{equation}

As an unit for numerical requirements, we take the BD field (inversely proportional to the Planck mass squared) equal to $1$ in present era ($M_P = 1/\sqrt{8\pi G} \simeq 2.43\times 10^{18}$ GeV) in reduced Planck mass unit. The BD scalar is $2$ dimensional and the Higgs is $1$ dimensional. $\omega$ and $\xi$ are dimensionless. The cosmological field equations are found to be (for detailed derivation see \citep{sola0, chakrabarti})

\begin{eqnarray}\nonumber
&& {{3}H^{2}\psi}+{{3}H{\dot{\psi}}}-\frac{\omega}{2}\frac{\dot{\psi}^{2}}{\psi} + U(\phi, \psi) -\frac{1}{2}\dot{\phi}^{2}-V(\phi) \\&&\nonumber
+ 6\xi H^{2}\phi^{2} + 12\xi H \dot{\phi}\phi - {9}\theta H^{2}\frac{\dot{\phi}^{2}}{\phi^{2}} -  6(\theta-\varsigma)\dot{H}\frac{\dot{\phi}^{2}}{\phi^{2}} \\&&
+ 6(\theta-\varsigma)H\frac{\dot{\phi}\ddot{\phi}}{\phi^{2}} - 6(\theta -\varsigma){H}\frac{\dot{\phi}^{3}}{\phi^{3}} = 0,
 \label{eq:EoM-metric}
\end{eqnarray}

\begin{eqnarray}\nonumber
&&\ddot{\phi}+3 H \dot{\phi}-12\xi\dot{H}\phi - 24\xi H^{2}\phi + \frac{d V}{d\phi} + 6 (2\theta - \varsigma) H^{2} \\&& \nonumber
\Bigg(\frac{\ddot{\phi}}{\phi^{2}} - \frac{\dot{\phi}^{2}}{\phi^{3}}\Bigg) + 18 (2\theta -\varsigma){H}^{3}\frac{\dot{\phi}}{\phi^{2}} + 6 (7\theta - 5\varsigma) H \dot{H}\frac{\dot{\phi}}{\phi^{2}} \\&&
+ 6 (\theta - \varsigma)\ddot{H}\frac{\dot{\phi}}{\phi^{2}} + 6(\theta -\varsigma)\dot{H}\Big(\frac{\ddot{\phi}}{\phi^{2}}-\frac{\dot{\phi}^{2}}{\phi^{3}}\Big) - \frac{\partial U}{\partial \phi} = 0, \label{eq:EoM-phi}
\end{eqnarray}

and

\begin{equation}
3\dot{H}+6{H}^{2} - \omega \frac{\ddot{\psi}}{\psi}+\frac{ \omega}{2} \frac{\dot{\psi}^{2}}{{\psi}^{2}}-3H\omega\frac{\dot{\psi}}{\psi}+ \frac{\partial U}{\partial \psi} = 0. \label{eq:EoM-psi}
\end{equation}

Even with the steep modifications in the lagrangian, $G^{eff}_{\mu\nu}$ satisfies covariant conservation. To use the results of the statefinder diagnostic and solve the modified field equations we use the ansatz 
\begin{eqnarray}
&&\frac{\partial U}{\partial \phi} \sim \phi_{0}, \\&&
\frac{\partial U}{\partial \psi} \sim \psi_{0},
\end{eqnarray}
with additional parameters $\phi_0$ and $\psi_0$. Although it looks like just a simplifying assumption, this special assumption is well-tested in the models of unification satisfying phenomenological constraints \citep{farrar, huey, micheletti}. In the context of one of the most popular models of Dark Energy-Dark Matter unification \citep{bertolamiinter} the interaction is taken as

\begin{equation}\label{crosspot1}
U(\psi,\phi) = e^{-\lambda \psi}P(\psi,\phi).
\end{equation} 

In comparison, the present ansatz essentialy leads to an additional constraint over $P(\psi,\phi)$ as
\begin{equation}\label{crosspot2}
P(\psi,\phi) = \frac{\phi_{0}}{\lambda}e^{\lambda\phi} + Q(\psi).
\end{equation} 

\begin{figure}
\begin{center}
\includegraphics[width=0.40\textwidth]{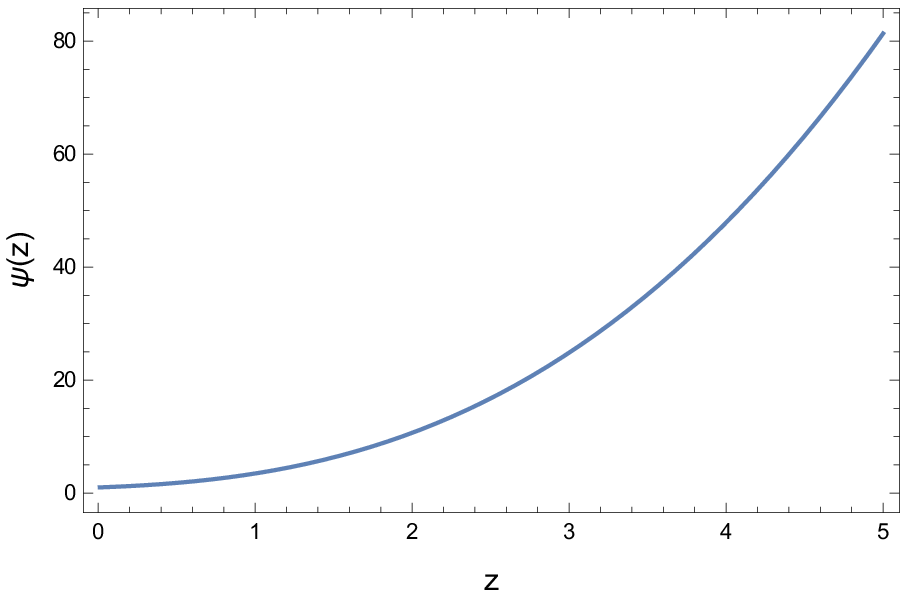}
\includegraphics[width=0.40\textwidth]{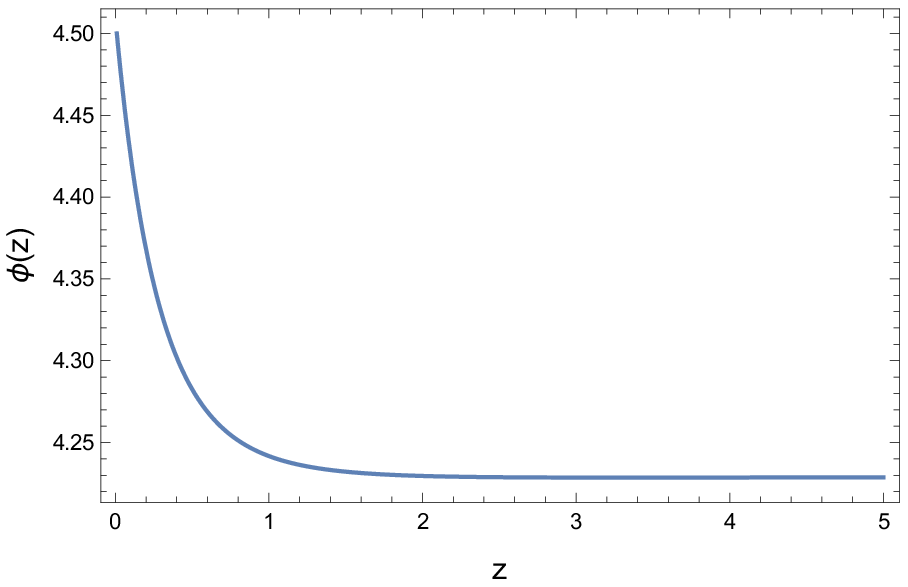}
\caption{Evolution of the Brans-Dicke field (top panel) and the Higgs field (bottom panel) at low redshift.}
\label{BDH1}
\end{center}
\end{figure}

It has already been proved that a late-time cosmology can be consistently described with a polynomial of $\psi$ in $P(\psi,\phi)$ \citep{bertolamiinter}. The present setup leads to the conclusion that a cosmological consistency can also be attained using an interaction function exponential in $\phi$. Using this, we finally transform Eq. (\ref{eq:EoM-psi}) into
\begin{eqnarray}\label{psibda}\nonumber
&& \psi^{\circ\circ} = \frac{{\psi^{\circ}}^2}{2\psi} + \frac{\psi^{\circ}}{a^2}\Big[(3+a^{-3.06})^{-\frac{1}{2}} a^{-\frac{23}{20}} \Big\lbrace (3+a^{-3.06})^{\frac{1}{2}} \\&&\nonumber
- \frac{3}{2}a^{-3.06}(3+a^{-3.06})^{-\frac{1}{2}} \Big\rbrace + \frac{1}{a} \Big] + \frac{\psi_{0}\psi}{\omega}a^{\frac{17}{10}}\\&&\nonumber
(3+a^{-3.06})^{-1} + \frac{\psi}{\omega} \Big[3(3+a^{-3.06})^{-\frac{1}{2}} a^{\frac{17}{10}} \Big\lbrace (3+a^{-3.06})^{\frac{1}{2}} \\&&
- \frac{3}{2}a^{-3.06}(3+a^{-3.06})^{-\frac{1}{2}} \Big\rbrace - 3a^2 \Big].
\end{eqnarray}

\begin{figure}
\begin{center}
\includegraphics[width=0.40\textwidth]{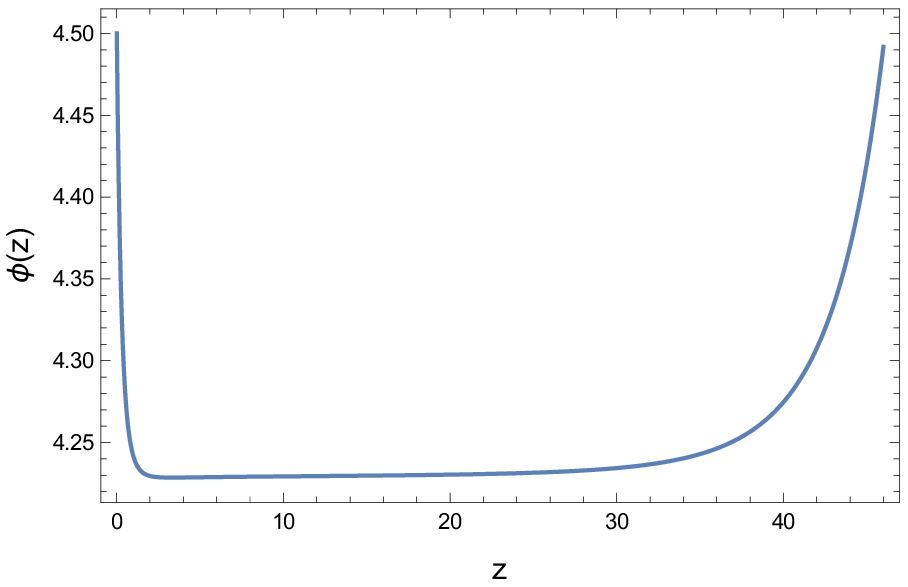}
\includegraphics[width=0.40\textwidth]{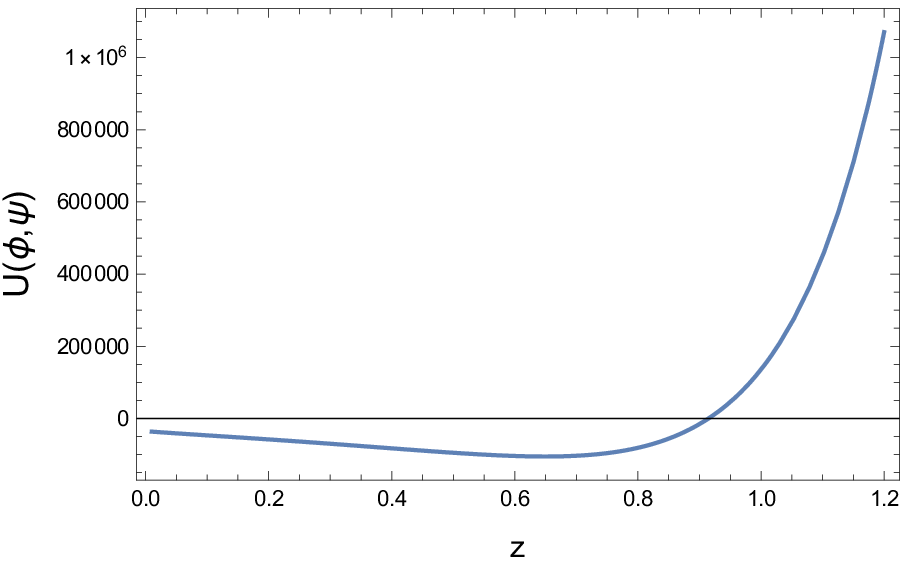}
\caption{Evolution of the Higgs field and the strength of the Brans-Dicke-Higgs interaction as a function of redshift.}
\label{BDH2}
\end{center}
\end{figure}

As usual, overhead circles denote redshift derivatives. Following a similar step by step transformation one can write the $\phi$ equation as well as the expression for the interaction from the independent field equations. For a precise presentation we give the numerical solutions straightaway. We choose the parameter $\psi_{0} \sim O(10^{-1})$ and the dimensionless $\omega \sim 10^5$. The evolving nature of the geometric field $\psi$ and the Higgs scalar $\phi$ are shown in Fig. \ref{BDH1}. Since the evolutions are in reduced Planck mass unit, $\psi_{(z = 0)} = 1 M_{P}^{2}$. We note that during late-time acceleration, the Higgs scalar with derivative interaction dominates the geometric BD scalar. This arguement is also true for any other phase of cosmic acceleration. One can find a hint of this from the plot of Higgs scalar for $z \geq 40$, given in the top panel of Fig. \ref{BDH2}. During deceleration ($z > 1$) Higgs varies very slowly with $z$ and the BD scalar dominates the dynamics. Thus, the geometric scalar can be associated with a cosmic deceleration and the interacting Higgs is connected with cosmic acceleration. Intriguingly, the interaction $U(\phi, \psi)$ acts as a switch between two scalars, their successive dominating nature and the continuity of transition between epochs, as shown in the lower panel of Fig. \ref{BDH2}. During late time acceleration, the interaction is negligible and effectively produces a non-minimally coupled scalar cosmology. On the other hand, a deceleration is influenced by a dominant interaction of the scalars as well as the geometric scalar, thus mimicking a strongly interacting Brans-Dicke theory \citep{chakrabarti}. The transition of the universe coincides with the sharp downfall of the interaction rond the redshift of transition, i.e., $z \sim 1$. 

\begin{figure}
\begin{center}
\includegraphics[width=0.40\textwidth]{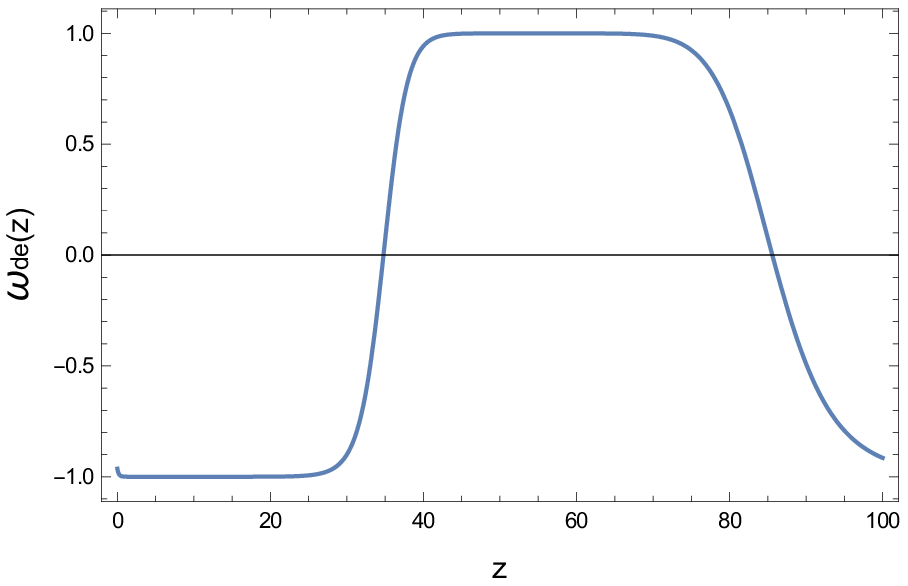}
\includegraphics[width=0.40\textwidth]{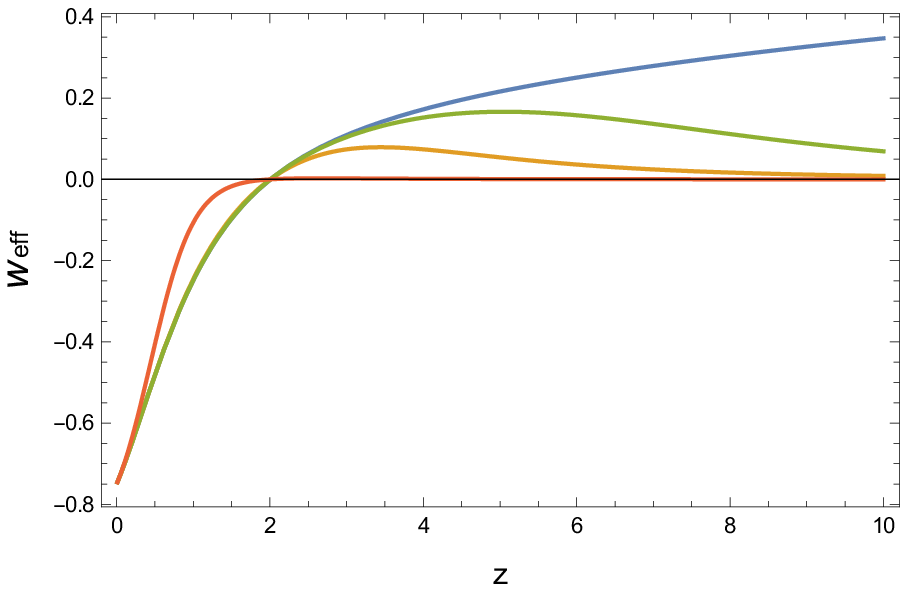}
\caption{Equation of State of the Higgs and the effective Equation of State of the system vs $z$. Blue curve $\rightarrow \omega = 50000$, Green curve $\rightarrow \omega = 60000$, Orange curve $\rightarrow \omega = 70000$ and Red curve $\rightarrow \omega = 100000$.}
\label{BDH3}
\end{center}
\end{figure}

In this scalar-dominated cosmological setup with no perfect fluid description it is very important to look into the EOS of the separate scalars as well as the effective EOS of the entire system. We use the field Eqs. (\ref{eq:EoM-metric}) and (\ref{eq:EoM-psi}) to write
\begin{eqnarray}\nonumber
&& \rho_{eff} = -\frac{1}{\psi} \Bigg[{{3}H{\dot{\psi}}}-\frac{\omega}{2}\frac{\dot{\psi}^{2}}{\psi} + U(\phi, \psi) -\frac{1}{2}\dot{\phi}^{2}-V(\phi) \\&&\nonumber
+ 6\xi H^{2}\phi^{2} + 12\xi H \dot{\phi}\phi - {9}\theta H^{2}\frac{\dot{\phi}^{2}}{\phi^{2}} -  6(\theta-\varsigma)\dot{H}\frac{\dot{\phi}^{2}}{\phi^{2}} \\&&
+ 6(\theta-\varsigma)H\frac{\dot{\phi}\ddot{\phi}}{\phi^{2}} - 6(\theta -\varsigma){H}\frac{\dot{\phi}^{3}}{\phi^{3}}\Bigg],
 \label{effdens}
\end{eqnarray}
and
\begin{equation}
p_{eff} = \frac{2}{3} \Bigg[6{H}^{2} - \omega \frac{\ddot{\psi}}{\psi}+\frac{ \omega}{2} \frac{\dot{\psi}^{2}}{{\psi}^{2}}-3H\omega\frac{\dot{\psi}}{\psi}+ \frac{\partial U}{\partial \psi}\Bigg] - \rho_{eff}. \label{effpress}
\end{equation}

Fig. \ref{BDH3} shows the evolution of EOS parameters for the Higgs (top panel) as well as the entire system (bottom panel) with redshift. Both of these graphs show a clear sign of negative pressure around $z \sim 0$ which is necessary to drive the late-time acceleration. The Higgs EOS shows hint of periodicity but apart from that no signature of a sinusoidal Higgs VEV on the system can be found. This becomes even ore clear with the plot of $w_{eff}$ vs $z$, which is quite sensitive on the Brans-Dicke parameter $\omega$. $w_{eff}$ is plotted for different values of the BD parameter $\omega$ and for all the cases, a Dark Energy dominated late time era is evident. However, nature of the system prior to the present acceleration depends on $\omega$. The system is either radiation dominated or dust dominated for higher redshift and a greater value of $\omega$ pushes the theory closer to the cosmology of standard GR. A generalized Brans-Dicke-Higgs cosmology was recently discussed in details with a non-varying Higgs VEV \citep{chakrabarti} and the qualitative outcomes of the present discussion carries remarkable similarity with the same. This makes us comment that a a time evolving Higgs VEV makes negligible effect on the cosmological scale in a scalar dominated theory of gravity. We also note that the Higgs EOS at higher redshift hints (although roughly) that it is also quite possible to describe the unified time history of the universe (i.e., early inflation-deceleration-late time acceleration) with some form of Higgs scalar dominated theory, albeit on a more proper scale.

\subsection{Unified Cosmic History}
Before concluding the manuscript, we discuss the case of a varying Higgs VEV in the context of a unified cosmic history as far as possible. It is evident that using just a parameteric statefinder reconstruction this is not possible since the JLA or the BAO data are mainly centred around the late-time era. This requires better observational data sets as well as advanced numerical schemes which are not within the scope of the present work. However, we try to give an idea of the unification through a simple yet interesting mathematical toy model. Ideally a complete theory of cosmology should describe the unified time history of the universe (inflation-deceleration-present acceleration) instead of providing patches which are at best an effective dynamics of a few specific epochs (although more popular, cosmologies of specific epochs do have their share of failures). Similar motivations have led to a few compelling examples \citep{elizalde, capouni} of reconstruction of cosmic time history. We construct the mathematics from a conjecture on the scale factor describing the expansion history (on some scale) at the outset. We take the Hubble and the scale factor as 

\begin{eqnarray}\label{1.13}
&& H(t) = H_{0}+\frac{H_{1}}{t^{n}} \\&&
a(t) = a_{0}\exp \Big[H_{0}t -\frac{H_{1}}{(n-1)t^{n-1}}\Big].
\end{eqnarray}

\begin{figure}
\begin{center}
\includegraphics[width=0.40\textwidth]{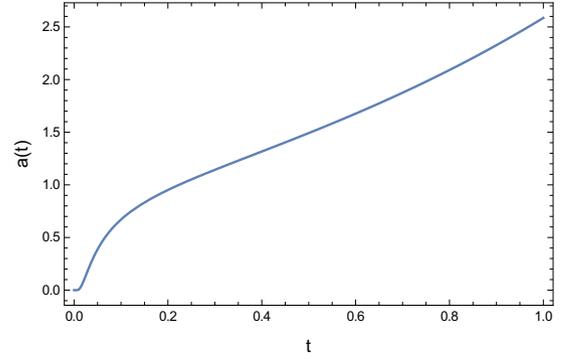}
\caption{Scale factor for a unified time history of the universe as a function of time.}
\label{unifiedplot}
\end{center}
\end{figure}

$H_{0}$, $H_{1}$, $n$ and $a_{0}$ are model parameters. We will work with positive time $t > 0$. Around $t \sim 0$ the scale factor goes to zero, marking the beginning of the universe. In Fig. \ref{unifiedplot}, we plot the scale factor as a function of cosmic time $t$, to establish an idea regarding the scale of the universe in this mathematical construct. We take the parameters $H_{0} = 1$, $H_{1} = 0.05$, $n = 4$ and $a_{0} = 1$. The initial acceleration is realized within $t \sim 0.05$ and subsequently epochs of deceleration and a second acceleration comes into play. \\

This is more clearly demonstrated using the evolving effective EOS

\begin{eqnarray}\label{1.14}
&& w_{\rm eff} = -1-\frac{2\dot{H}}{{3H}^{2}}, \\&&
=-1+\frac{2nH_{1}t^{n-1}}{\Bigg(H_{0}t^{n}+H_{1}\Bigg) ^{2}}.
\end{eqnarray}

We note that in $t \to 0$ limit as well as for large $t$, the effective EOS goes to $-1$. Therefore we have two epochs of accelerations, an early and a late time acceleration. There are two critical points of phase transition in the time history of the universe. This is realized from the behavior of $\ddot a/a$

\begin{equation}
\frac{\ddot a}{a}=\dot H +H^{2}=-\frac{nH_{1}}{t^{n+1}} + \left( H_{0}+\frac{H_{1}}{t^{n}}\right) ^{2}\ . \label{1.15}
\end{equation}

Provided $\frac{4H_{0}}{n} \leq 1$, these critical phase transitions are realized at the zeros of $\ddot{a}/a$ realized at
\begin{equation}
t_{\pm} \approx \left[ \sqrt{nH_{1}} \,\, \frac{\left( 1\pm
\sqrt{1-\frac{4H_{0} }{n}} \, \right)}{2H_{0}} \, \right]^{2/n}.
\label{1.16}
\end{equation}
Therefore the entire time history is compiled as
\begin{itemize}
\item {Time domain $0 < t < t_{-}$ is the early inflation.}
\item {Time domain $t_{-} < t < t_{+}$ is the era of deceleration.}
\item {Time domain $t > t_{+}$ is the late-time or the present acceleration.}
\end{itemize}

\begin{figure}
\begin{center}
\includegraphics[width=0.40\textwidth]{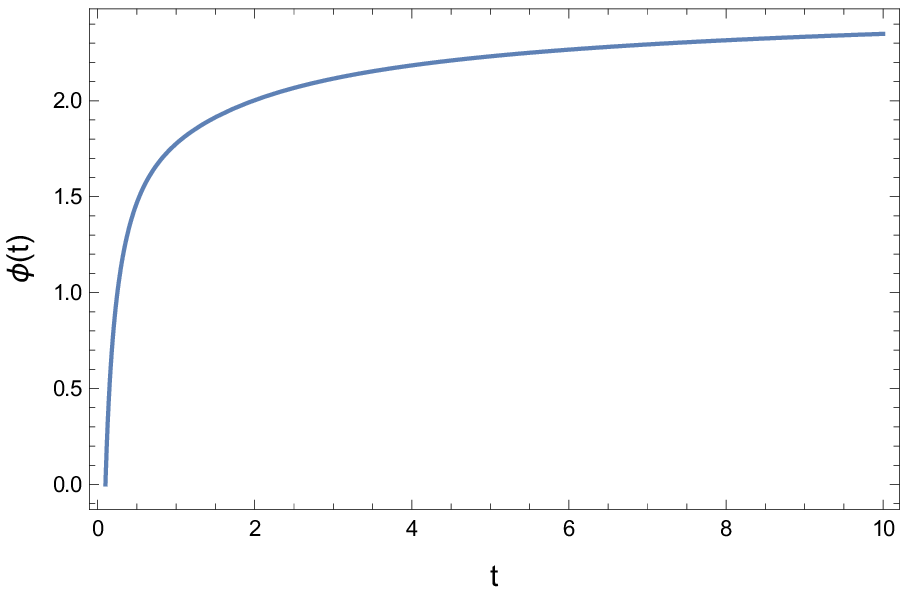}
\includegraphics[width=0.40\textwidth]{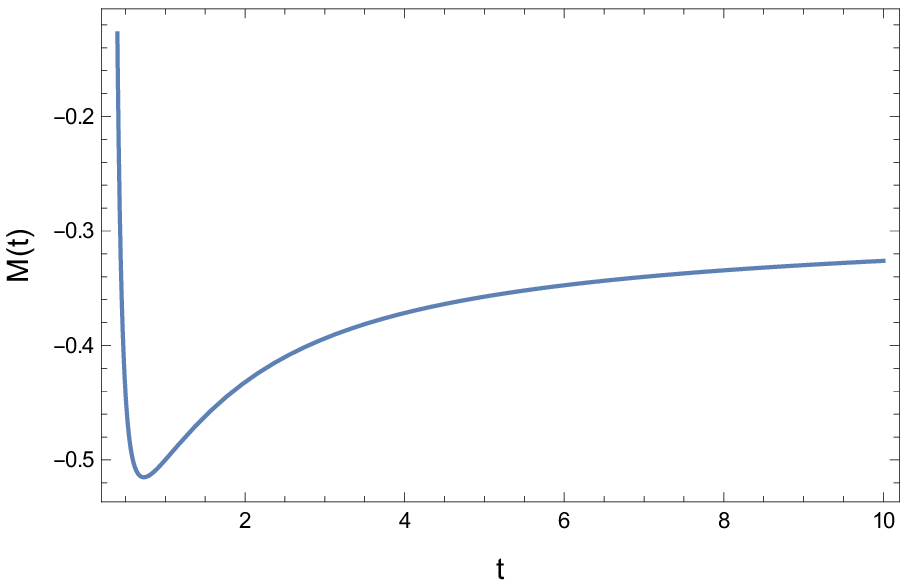}
\includegraphics[width=0.40\textwidth]{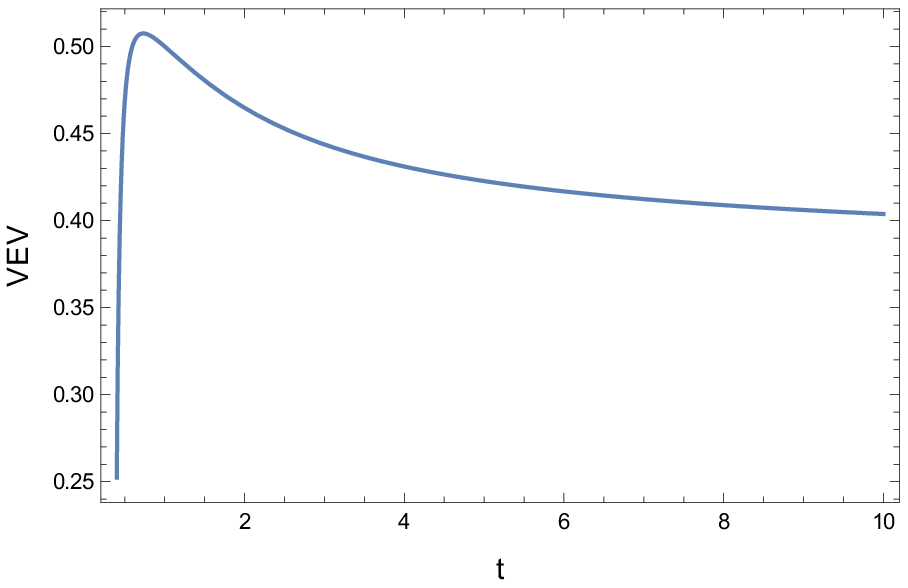}
\caption{Evolution of the scalar field, $M(t)$ and the Higgs VEV for minimally coupled scalar field theory depicting a unified time history of the universe as a function of time.}
\label{unifiedplot_min}
\end{center}
\end{figure}

Overall this defines the conjecture over the cosmic time history. The scalar energy momentum distribution supporting this evolution must be determined from the field equations of the theory under consideration. In all relevant cases we choose the self interaction to be of Higgs nature as in Eq. (\ref{potential}), i.e., we treat the coefficiet of $\phi^2$ as a time evolving independent function. The first theory we look into is a simple minimally coupled scalar theory as discussed in subsection $3.1$. The field equtions are
\begin{equation} \label{fe1minimaluni}
3\Big(\frac{\dot{a}}{a}\Big)^{2} = \rho_{m} + \rho_{\phi} = \rho_{m} + \frac{\dot{\phi}^{2}}{2} + V\left( \phi \right),
\end{equation} 
 
\begin{equation} \label{fe2minimaluni}
-2\frac{\ddot{a}}{a}-\Big(\frac{\dot{a}}{a}\Big)^{2} = p_{m} + p_{\phi} = p_{m} + \frac{\dot{\phi}^{2}}{2}-V\left( \phi \right),
\end{equation}

\begin{equation} \label{phiminimaluni}
\ddot{\phi} + 3\frac{\dot{a}}{a}\dot{\phi} + \frac{dV(\phi)}{d\phi} = 0.  
\end{equation}     

alongwith the interaction potential 
\begin{equation}\label{potentialuni}
V(\phi) = V_{0} + M(t) \phi^{2} + \frac{\lambda}{4} \phi^{4}.
\end{equation}

We assume that the fluid distribution accompanying the time evolving scalar field is pressureless and this allows us the freedom to write the fluid density as proportional to $\frac{1}{a^3}$. We solve the set of equations numerically using Eq. (\ref{1.13}) and plot the scalar field, $M(t)$ and the VEV as a function of cosmic time in Fig. \ref{unifiedplot_min}. The graph on top of the figure shows that the scalar field becomes increasingly dominant with cosmic time and asymptotically reaches a constant value as the present acceleration evolves into future. We recall that from the self-interaction potential the Higgs VEV $\nu$ can be calculated as
\begin{eqnarray}
&& \frac{\partial V}{\partial \phi} \bigg\rvert_{\nu} = 0,
\nu = \sqrt{\frac{-2M(t)}{\lambda}}. 
\end{eqnarray}
As we can see from the middle and the bottom graph of Fig. \ref{unifiedplot_min}, the evolution of $M(t)$ as well as the VEV shows a clear sign of variation with cosmic time. The variation is quite dominant in an earlier epoch but becomes comparatively mild once a late-time acceleration starts (around $t \sim 1$ in the present scale).   \\

\begin{figure}
\begin{center}
\includegraphics[width=0.40\textwidth]{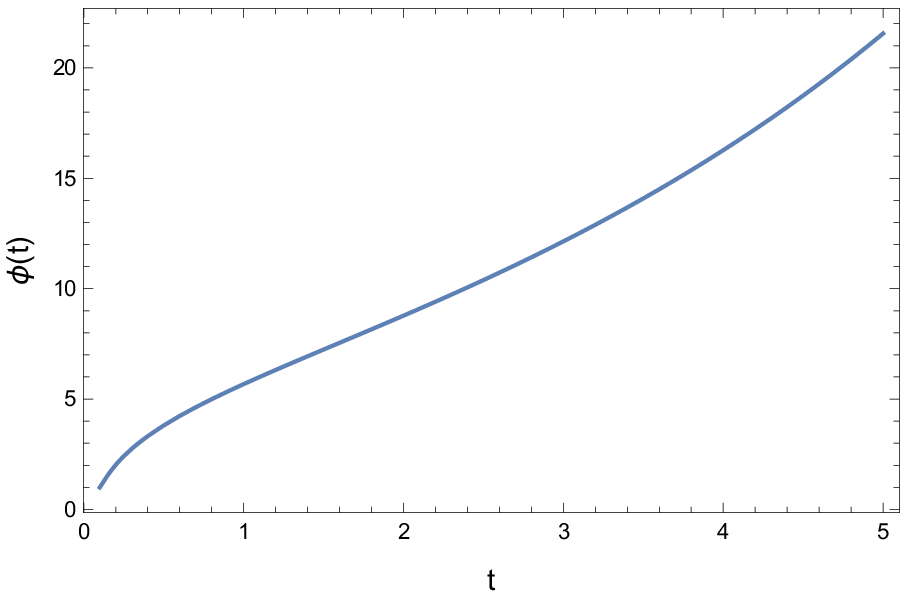}
\includegraphics[width=0.40\textwidth]{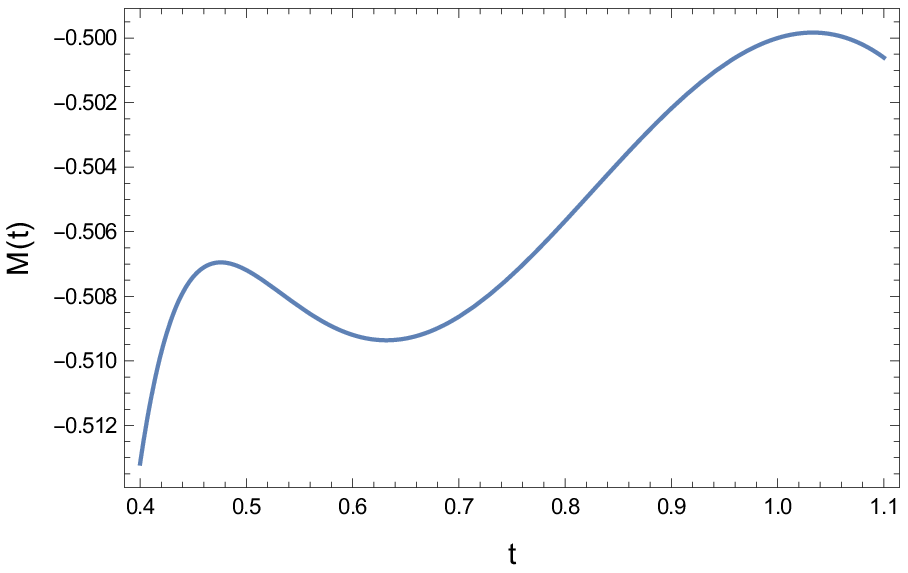}
\includegraphics[width=0.40\textwidth]{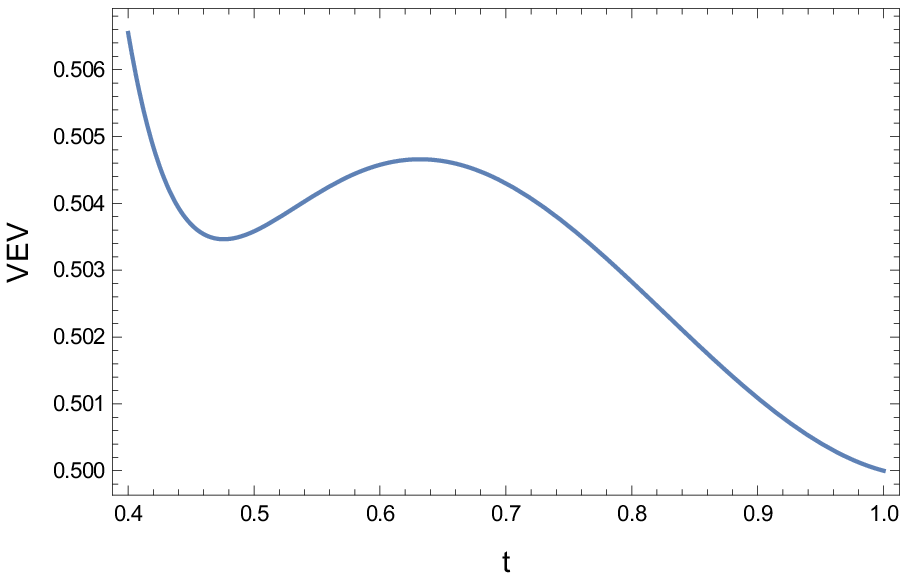}
\caption{Evolution of the scalar field, $M(t)$ and the Higgs VEV for non-minimally coupled scalar field theory depicting a unified time history of the universe as a function of time.}
\label{unifiedplot_nonmin}
\end{center}
\end{figure}

A similar mathematical excercise can be done using the set of field equations for a non-minimally coupled scalar and this leads to more intriguing hints. The field equations in this case are
\begin{equation}
6 W H^{2} + 6 W'H\dot{\phi} = \rho_{m} + \frac{1}{2}\dot{\phi}^2 + V(\phi),
\label{nonminfrw1uni}
\end{equation}

\begin{equation}
4W\dot{H} + 6WH^2 + 4W'H\dot{\phi} + 2W''\dot{\phi}^2 + 2W'\ddot{\phi} = -p_{m} -\frac{\dot{\phi}^2}{2} + V(\phi),
\label{nonminimalfrw2uni}
\end{equation}

\begin{equation}
\ddot{\phi} + 3H\dot{\phi} - 6W'\left[\dot{H} + H^{2}\right] + \frac{dV}{d\phi} = 0.
\label{nonminKGiniuni}
\end{equation}

We once again take a quadratic scalar-geometry interaction as 
\begin{equation}
W(\phi) = \frac{1}{2}(1 + U_{0}\phi^2).
\end{equation}

Once again, we take the fluid distribution accompanying the time evolving scalar field to be pressureless so that the fluid density can be replaced in terms of scale factor. We solve the set of equations and plot the scalar field, $M(t)$ and the VEV as a function of time in Fig. \ref{unifiedplot_nonmin}. The graph on top shows the nature of the evolving scalar field. The plot of $M(t)$ and the VEV is given in the middle and the bottom panel and shows a clear sign of variation with cosmic time. More intriguingly, their evolution suggests of a nascent periodicity in the Higgs VEV. Although as a function of cosmic time this can at most serve as a toy model, the hints from this can be quite interesting. It is not impossible to find inspiration to look into the non-trivial periodicity in the proton-to-electron mass ratio on a proper scale and in terms of cosmic redshift. While a minimal scalar interaction did not unearth the oscillatory nature of the proton-to-electron mass ratio, a non-minimal scalar interaction did expose a glimpse. We conclude this section with the comment that a more generalized scalar extended theory of gravity may support a more prominent oscillating behavior of the proton to electron mass ratio or the Higgs VEV. At the same time it can describe a unified cosmic time history maintaining the observational consistency. Such a theory has a strong onus to become `the' better theory of gravity and will be addressed elsewhere in near future. 

\section{Conclusion}
The present manuscript sheds some light on the existence of unorthodox physics at the confluence of classical gravity and interacting fundamental particles. It describes an evolving nature of an otherwise constant entity, the proton-to-electron mass ratio. The ratio is connected to the vacuum expectation value of an interacting Higgs scalar field. We discuss the role of this variation in the context of late-time cosmology. The claim of an evolving Higgs VEV or it's viability in different cosmic epochs has intrigued relativists as well as particle physicists for some time and has produced enough evidence to be taken into account while considering generalized scalar extended theories of gravity. \\

From this purview, the present attempt is a simple mathematical excercise to describe a varying Higgs VEV during a smooth transition of the universe between the matter/radiation dominated deceleration and the dark energy dominated late time acceleration. We show that it is quite possible to support the reasonable transition with a mild sinusoidal variation of the Higgs VEV. The variation is incorporated in the theory of gravity using a specific form of the cosmological Higgs self-interaction potential, as in Eq. (\ref{potential}). A sinusoidal variation is a bit non-trivial, but matches well with the observational data of the variation from molecular absorption spectra of a series of Quasars. We neither claim that this is the only possible functional form of the variation to fit in with the data, nor is it the only possible way to theoretically describe the variation. It is just a possibility that ticks all the boxes. Moreover, this formalism is simple enough to employ a statefinder reconstruction of the late-time cosmology and reassemble the pieces of the theory supporting all of the claims. The reconstructon is analytical and enables us to construct the viable cosmic evolution from a kinematic quantity alone. This simplicity allows us to discuss not just the case of a simple minimally coupled Higgs scalar, but further generalized cases as well, such as the case of a derivative Higgs interaction.  \\

In all the relevant cases, we particularly focus on the evolving equation of state of the constituent elements of the universe, such as one or more scalars as well as a perfect fluid description in some cases. If the Higgs VEV must be oscillatory, we find that the EOS of the scalar dark energy should exhibit some oscillation in the low redshift regime. More importantly, this evolution leaves almost no mark on the effective EOS of the total system which simply is the standard dark energy dominated universe at present era and a dust dominated universe preceding the same. We also find the generic nature of the perfect fluid EOS which shows a dominant periodicity, atleast around the redshift of transition from deceleration into acceleration. It is possible that the oscillating behavior of the EOS of the scalar and the fluid cancels each one out to stabilize the system. Although not under the scope of just one work, one can imagine that in these setups, the interacting scalar is supposed to play the role of a Dark Energy fluid and the fluid can play the role of a Dark Matter. For instance, for a higher $z$ the fluid EOS goes to zero, pointing towards a dust-like behavior of the fluid. It is more likely to find the signatures of a varying Higgs VEV or the varying proton-to-electron mass ratio in a better description of the Dark Matter component, such as an Axion scalar dark matter distribution \citep{sikivie, arvanitaki}. In this regard we also demonstrate a possible interacting Dark Matter-Dark Energy scenario with a generalized two-scalar theory with one geometric scalar and the Higgs scalar with non-derivative as well as a derivative interaction. In this case, the interaction between the two scalars acts as a conductor of the smooth transition of the universe across different epochs. \\

Before conclusion, we include a simple toy model of unified cosmic history and discuss the role of a varying Higgs VEV in driving the time evolution. We show that to describe the cosmic evolution as in an inflation-deceleration-acceleration succession (as in the conjecture Eq. (\ref{1.13}), the Higgs VEV must exhibit signs of variation with cosmic time, leading to a non-trivial variation of the proton-to-electron mass ratio. We verify this with a minimally coupled scalar as well as a non-minimally coupled scalar model. The periodicity in the VEV becomes prominent when the scalar interacts with curvature. A pressureless dust matter accompanies the Higgs scalar throughout the cosmic history without which the solution for the Higgs or the VEV becomes inconsistent. Although just a toy model, this shows us the possibility that physics beyond our usual understandings can easily hide behind the expected large scale evolutions of our known universe. Through a combination of bold theoretical considerations as well as new fronts of observational astronomy such as the Quasar spectra in this case, one can continue attempting to unearth these facts.
\\

{\bf Data Availability Statement} This manuscript has no associated data or the data will not be deposited.

\bibliographystyle{unsrt}

\begin{thebibliography}{}

\bibitem[\protect\citeauthoryear{Ade et. al.}{2014}]{planck}
Ade P. A. R. et. al., Planck collaboration, 2014, Astron. Astrophys. 571, A16.

\bibitem[\protect\citeauthoryear{Alam et. al.}{2003}]{alam}
Alam U., Sahni V., Saini T. and Starobinsky A. A., 2003, Mon. Not. Roy. Astron. Soc. 344 : 1057.

\bibitem[\protect\citeauthoryear{Alexander, Barrow and Magueijo}{2016}]{alexander}
Alexander S., Barrow J. D. and Magueijo J., 2016, Class. Quantum Gravity, 33, 14LT01.

\bibitem[\protect\citeauthoryear{Amendola}{1993}]{amendola0}
Amendola L., 1993, Phys. Lett. B, 301, 175.

\bibitem[\protect\citeauthoryear{Amendola}{1999}]{amendola00}
Amendola L., 1999, Phys. Rev. D., 60, 043501.

\bibitem[\protect\citeauthoryear{Amendola et. al.}{2012}]{amendola}
Amendola L., Leite  A., Martins C., Nunes  N., Pedrosa P. and Seganti A., 2012, Phys.
Rev. D 86, 063515.

\bibitem[\protect\citeauthoryear{Anchordoqui and Goldberg}{2003}]{anchor}
Anchordoqui L. and Goldberg H., 2003, Phys. Rev. D. 68, 083513.

\bibitem[\protect\citeauthoryear{Anderson et al.}{2012}]{bossanderson}
Anderson L. et al. (BOSS collaboration), 2012, Mon. Not. Roy. Astron. Soc. 441, 24.

\bibitem[\protect\citeauthoryear{Arvanitaki et al.}{2010}]{arvanitaki}
Arvanitaki A., Dimopoulos S., Dubovsky S., Kaloper N. and March-Russell J., 2010, Phys. Rev. D. 81, 123530.

\bibitem[\protect\citeauthoryear{Atkins and Calmet}{2013}]{atkins}
Atkins M. and Calmet X., 2013, Phys. Rev. Lett. 110, 051301.

\bibitem[\protect\citeauthoryear{Avelino, Martins and Oliveira}{2004}]{avelino}
Avelino P. P., Martins C. J. A. P. and Oliveira J. C. R. E., 2004, Phys. Rev. D. 70, 083506.

\bibitem[\protect\citeauthoryear{Avelino et. al.}{2006}]{avelino0}
Avelino P. P., Martins C. J. A. P., Nunes N. J. and Olive K. A., 2006, Phys. Rev. D. 74, 083508.

\bibitem[\protect\citeauthoryear{Avelino}{2008}]{avelino1}
Avelino P. P., 2008, Phys. Rev. D. 78, 043516.

\bibitem[\protect\citeauthoryear{Bagdonaite et al.}{2012}]{bagdon}
Bagdonaite J., Murphy M. T., Kaper L. and Ubachs W., 2012, Mon. Not. Roy. Astron. Soc. 421, 419.

\bibitem[\protect\citeauthoryear{Bagdonaite et al.}{2014}]{bagdon0}
Bagdonaite J., Salumbides E. J., Preval S. P., Barstow M. A., Barrow J. D., Murphy  M. T. et al., 2014, Phys. Rev. Lett. 113, 123002.

\bibitem[\protect\citeauthoryear{Bak and Rey}{2000}]{bakrey}
Bak D. and Rey S. J., 2000, Class. Quant. Gravit., 17, L83.

\bibitem[\protect\citeauthoryear{Banerjee and Pavon}{2001}]{nbpavon}
Banerjee N. and Pavon D., 2001, Class. Quantum Gravity, 18, 593.

\bibitem[\protect\citeauthoryear{Banerjee and Sen}{1997}]{nbsudipta}
Banerjee N. and Sen S., 1997, Phys. Rev. D, 56, 1334.

\bibitem[\protect\citeauthoryear{Barker}{1978}]{barker}
Barker B. M., 1978, ApJ, 219, 5.

\bibitem[\protect\citeauthoryear{Barrow and Magueijo}{2005}]{barrow}
Barrow J. D. and Magueijo J., 2005, Phys. Rev. D. 72, 043521.

\bibitem[\protect\citeauthoryear{Bento, Bertolami and Sen}{2002}]{bento0}
Bento M., Bertolami O. and Sen A., 2002, Phys. Rev. D, 66, 043507.

\bibitem[\protect\citeauthoryear{Bento, Bertolami and Santos}{2004}]{bento}
Bento M. d. C., Bertolami O. and Santos N. M. C., 2004, Phys. Rev. D. 70, 107304.

\bibitem[\protect\citeauthoryear{Bergmann}{2004}]{bergmann}
Bergmann P. G., 1968, Int. J. Theor. Phys., 1, 25.

\bibitem[\protect\citeauthoryear{Bertolami and Martins}{2000}]{bertolami}
Bertolami O. and Martins P., 2000, Phys. Rev. D., 61, 064007.

\bibitem[\protect\citeauthoryear{Bertolami et. al.}{2012}]{bertolamiinter}
Bertolami O., Carrilho P. and P\'aramos J, 2012, Phys. Rev. D 86, 103522.

\bibitem[\protect\citeauthoryear{Betoule et al.}{2014}]{jla}
Betoule M. et al., 2014, Astron. Astrophys. 568, A22.

\bibitem[\protect\citeauthoryear{Beutler et. al.}{2011}]{6dF}
Beutler F. et. al., 2011, Mon. Not. Roy. Astron. Soc. 416, 3017.

\bibitem[\protect\citeauthoryear{Bezrukov and Shaposhnikov}{2008}]{bezrukov}
Bezrukov F. and Shaposhnikov M., 2008, Phys. Lett. B., 659, 703.

\bibitem[\protect\citeauthoryear{Bilic, Tupper and Viollier}{2002}]{bilic}
Bilic N., Tupper G. B. and Viollier R. D., 2002, Phys. Lett. B, 535, 17.

\bibitem[\protect\citeauthoryear{Birrell and Davies}{1980}]{birrell}
Birrell N. D. and Davies P. C. W., 1980, Phys. Rev. D. 22, 322.

\bibitem[\protect\citeauthoryear{Blake et. al.}{2012}]{ohd5}
Blake C. et. al., 2012, Mon. Not. Roy. Astron. Soc. 425, 405.

\bibitem[\protect\citeauthoryear{Brans and Dicke}{1961}]{bransdicke}
Brans C. H. and Dicke R. H., 1961, Phys. Rev., 124, 925.

\bibitem[\protect\citeauthoryear{Callan, Jr., Coleman and Jackiw}{1970}]{callan}
Callan Jr. C. G., Coleman S. R. and Jackiw R., 1970, Annal. Phys. 59, 42.

\bibitem[\protect\citeauthoryear{Calmet and Fritzsch}{2002}]{calmet0}
Calmet X. and Fritzsch H., 2002, Phys. Lett. B. 540, 173.

\bibitem[\protect\citeauthoryear{Calmet and Fritzsch}{2006}]{calmet1}
Calmet X. and Fritzsch H., 2006, Europhys. Lett. 76, 1064.

\bibitem[\protect\citeauthoryear{Calmet and Keller}{2015}]{calmet}
Calmet X. and Keller M., 2015, Mod. Phys. Lett. A30, 1540028.

\bibitem[\protect\citeauthoryear{Calmet}{2017}]{calmet00}
Calmet X., 2017, Eur. Phys. J. C. 77, 729.

\bibitem[\protect\citeauthoryear{Campbell and Olive}{1995}]{campbell}
Campbell B. A. and Olive K. A., 1995, Phys. Lett. B. 345, 429.

\bibitem[\protect\citeauthoryear{Capozziello and Lambiase}{1999}]{capolamb}
Capozziello S. and Lambiase G., 1999, Gen. Relativ. Gravit., 31, 1005.

\bibitem[\protect\citeauthoryear{Capozziello, Nojiri and Odintsov}{2006}]{capouni}
Capozziello S., Nojiri S. and Odintsov S. D., 2006, Phys. Lett. B. 632, 597.

\bibitem[\protect\citeauthoryear{Carroll}{1998}]{carroll}
Carroll S. M., 1998, Phys. Rev. Lett. 81, 3067.

\bibitem[\protect\citeauthoryear{Casadio, Iafelice and Vacca}{2007}]{casadio}
Casadio R., Iafelice P. L. and Vacca  G. P., 2007, Nucl. Phys. B. 783 : 1.

\bibitem[\protect\citeauthoryear{Chakrabarti}{2021}]{chakrabarti}
Chakrabarti S., 2021, MNRAS 502, 1895.

\bibitem[\protect\citeauthoryear{Chamoun, Landau and Vucetich}{2001}]{chamoun}
Chamoun N., Landau S. J. and Vucetich H., 2001, Phys. Lett. B. 504, 1.

\bibitem[\protect\citeauthoryear{Chand et. al.}{2004}]{chand}
Chand H., Srianand R., Petitjean P. and Aracil B., 2004, Astron. Astrophys. 417, 853 ; Phys. Rev. Lett. 92, 121302.

\bibitem[\protect\citeauthoryear{Chen et. al.}{2016}]{chen}
Chen Y., Ratra B., Biesiada M., Li S. and Zhu Z. H., 2016, Astrophys. J. 829, 61.

\bibitem[\protect\citeauthoryear{Chernikov and Tagirov}{1968}]{chernikov}
Chernikov N. A. and Tagirov E. A., 1968, Annales Poincare Phys. Theor. A9, 109. 

\bibitem[\protect\citeauthoryear{Chiba et al.}{2007}]{chiba0}
Chiba T., Kobayashi T., Yamaguchi M. and Yokoyama J., 2007, Phys. Rev. D. 75, 043516.

\bibitem[\protect\citeauthoryear{Chiba}{2011}]{chiba}
Chiba T., 2011, Prog. Theor. Phys. 126, 993.

\bibitem[\protect\citeauthoryear{Chuang and Wang}{2013}]{ohd3}
Chuang C. H. and Wang Y., 2013, Mon. Not. Roy. Astron. Soc. 435, 255.

\bibitem[\protect\citeauthoryear{Clarkson and Zunckel}{2010}]{clarkson}
Clarkson C. and Zunckel C., 2010, Phys. Rev. Lett. 104, 211301.

\bibitem[\protect\citeauthoryear{Clifton and Barrow}{2006}]{clifton}
Clifton T. and Barrow J. D., 2006, Phys. Rev. D, 73, 104022.

\bibitem[\protect\citeauthoryear{Copeland, Nunes and Pospelov}{2004}]{copeland1}
Copeland E. J., Nunes N. J. and Pospelov M., 2004, Phys. Rev. D. 69, 023501.

\bibitem[\protect\citeauthoryear{Copeland, Sami and Tsujikawa}{2006}]{copeland}
Copeland E. J., Sami M. and Tsujikawa S., 2006, Int. J. Mod. Phys. D. 15 : 1753.

\bibitem[\protect\citeauthoryear{Crittenden et. al.}{2009}]{critt}
Crittenden R., Pogosian L. and Zhao G., 2009, JCAP 0912, 025.

\bibitem[\protect\citeauthoryear{Cruz Perez and Sol\`a}{2018}]{sola2}
Cruz P\'erez J. and Sol\`a J., 2018, Mod. Phys. Lett. A33, 1850228.

\bibitem[\protect\citeauthoryear{Damour}{2012}]{damour}
Damour T., Class. Quant. Grav. 29 (2012) 184001.

\bibitem[\protect\citeauthoryear{Dapra et al.}{2017}]{dapra0}
Dapra M., Van der Laan M., Murphy M. T. and Ubachs W., 2017, Mon. Not. Roy. Astron. Soc. 465, 4057.

\bibitem[\protect\citeauthoryear{Dapra et al.}{2017}]{dapra}
Dapra M., Noterdaeme P., Vonk M., Murphy M. T. and Ubachs W., 2017, Mon. Not. Roy. Astron. Soc. 467, 3848.

\bibitem[\protect\citeauthoryear{Delubac et. al.}{2015}]{delu}
Delubac T. et. al., 2015, Astron. Astrophys. 574, A59.

\bibitem[\protect\citeauthoryear{Dent}{2007}]{dent}
Dent T., 2007, JCAP 0701, 013.

\bibitem[\protect\citeauthoryear{Dine et. al.}{2003}]{dine}
Dine M., Nir Y., Raz G. and Volansky T., 2003, Phys. Rev. D. 67, 015009.

\bibitem[\protect\citeauthoryear{Dirac}{1937}]{dirac1}
Dirac P. A. M., 1937, Nature 139, 323.

\bibitem[\protect\citeauthoryear{Dirac}{1938}]{dirac2}
Dirac P. A. M., 1938, Proc. Roy. Soc. Lond. A165, 199.

\bibitem[\protect\citeauthoryear{Donoghue}{1994}]{dono}
Donoghue J. F., 1994, Phys. Rev. D. 50, 3874. 

\bibitem[\protect\citeauthoryear{Doran}{2005}]{doran}
Doran M., 2005, JCAP 0504, 016.

\bibitem[\protect\citeauthoryear{Eisenstein et al.}{2005}]{eisen}
Eisenstein D. J. et. al., 2005, Astrophys. J. 633, 560.

\bibitem[\protect\citeauthoryear{Elizalde et al.}{2008}]{elizalde}
Elizalde E., Nojiri S., Odintsov S. D., Saez-Gomez D. and Faraoni V., 2008, Phys. Rev. D. 77 : 106005.

\bibitem[\protect\citeauthoryear{Faraoni}{1999}]{faraoni}
Faraoni V., 1999, Phys. Rev. D, 59, 084021.

\bibitem[\protect\citeauthoryear{Farooq and Ratra}{2013}]{farooq}
Farooq O. and Ratra B., 2013, ApJ, 766, L7.

\bibitem[\protect\citeauthoryear{Farrar and Peebles}{2004}]{farrar}
Farrar G. R. and Peebles P. J. E., 2004, ApJ, 604, 1.

\bibitem[\protect\citeauthoryear{Flambaum et. al.}{2007}]{flambaum}
Flambaum V. V., Kozlov  M. G. and Kozlov M. G., 2007, Phys. Rev. Lett. 98, 240801.

\bibitem[\protect\citeauthoryear{Ford}{1987}]{ford}
Ford L., 1987, Phys. Rev. D. 35, 2955.

\bibitem[\protect\citeauthoryear{Foreman-Mackey et. al.}{2013}]{foreman}
Foreman-Mackey D., Hogg D. W., Lang D. and Goodman J., 2013, PASP, 125, 306.

\bibitem[\protect\citeauthoryear{Fritzsch, Sola and Nunes}{2017}]{fritzsch}
Fritzsch H., Sol\`a Peracaula J. and Nunes R. C., 2017, Eur. Phys. J. C. 77, 193.

\bibitem[\protect\citeauthoryear{Gasser and Leutwyler}{1982}]{gasser}
Gasser J. and Leutwyler H., 1982, Phys. Rept. 87, 77.

\bibitem[\protect\citeauthoryear{Germani and Kehagias}{2010}]{germani}
Germani C. and Kehagias A., 2010, Phys. Rev. Lett., 105, 011302.

\bibitem[\protect\citeauthoryear{Gibbons and Hawking}{1977}]{gibbons}
Gibbons G. W. and Hawking S. W., 1977, Phys. Rev. D, 15, 2738.

\bibitem[\protect\citeauthoryear{Holden and Wands}{1998}]{holden}
Holden D. J. and Wands D., 1998, Class. Quantum Gravity, 15, 3271.

\bibitem[\protect\citeauthoryear{Holsclaw et. al.}{2010}]{hols}
Holsclaw T., Alam U., Sans\'o B., Lee H., Heitmann K., Habib S. and Higdon D., 2010, Phys. Rev. Lett. 105,
241302 ; Phys. Rev. D 82, 103502.

\bibitem[\protect\citeauthoryear{Huey and Wandelt}{2006}]{huey}
Huey and Wandelt, 2006, Phys. Rev. D, 74, 023519.

\bibitem[\protect\citeauthoryear{Huntemann et al.}{2014}]{huntemann}
Huntemann N., Lipphardt B., Tamm C., Gerginov V., Weyers S. and Peik E., 2014, Phys. Rev. Lett. 113, 210802.

\bibitem[\protect\citeauthoryear{Ishida and De Souza}{2011}]{ishida}
Ishida E. E. O. and De Souza R. S., 2011, Astron. Astrophys. 527, A49.

\bibitem[\protect\citeauthoryear{Ivanchik et al.}{2005}]{ivan}
Ivanchik A. et al., 2005, Astron. Astrophys. 440, 45.

\bibitem[\protect\citeauthoryear{Jacobson}{1995}]{jacobson}
Jacobson T., 1995, Phys. Rev. Lett., 75, 1260.

\bibitem[\protect\citeauthoryear{Jamil, Saridakis and Setare}{2010}]{jamil}
Jamil M., Saridakis E. N. and Setare M. R., 2010, J. Cosmol. Astropart. Phys., 1011, 032.

\bibitem[\protect\citeauthoryear{Ji}{1995}]{ji}
Ji X. D., 1995, Phys. Rev. Lett. 74, 1071.

\bibitem[\protect\citeauthoryear{Kamenshchik, Moschella and Pasquier}{2001}]{kamen}
Kamenshchik A. Y., Moschella U. and Pasquier V., 2001, Phys. Lett. B, 511, 265.

\bibitem[\protect\citeauthoryear{Kanekar}{2011}]{kanekar}
Kanekar N., 2011, Astrophys. J. 728, L12.

\bibitem[\protect\citeauthoryear{King et al.}{2008}]{king0}
King J. A., Webb J. K., Murphy M. T. and Carswell R. F., 2008, Phys. Rev. Lett. 101, 251304.

\bibitem[\protect\citeauthoryear{King et al.}{2011}]{king}
King J. A., Murphy M. T., Ubachs W. and Webb J. K., 2011, Mon. Not. Roy. Astron. Soc. 417, 3010.

\bibitem[\protect\citeauthoryear{La and Steinhardt}{1989}]{la}
La D. and Steinhardt P., 1989, Phys. Rev. Lett., 62, 376.

\bibitem[\protect\citeauthoryear{Landau and Vucetich}{2002}]{landau}
Landau S. J. and Vucetich H., 2002, Astrophys. J. 570, 463.

\bibitem[\protect\citeauthoryear{Langacker, Segre and Strassler}{2002}]{lang}
Langacker P., Segre G. and Strassler M. J., 2002, Phys. Lett. B. 528, 121.

\bibitem[\protect\citeauthoryear{Lee, Olive and Pospelov}{2004}]{lee}
Lee S., Olive K. and Pospelov M., 2004, Phys. Rev. D. 70, 083503.

\bibitem[\protect\citeauthoryear{Lee}{2007}]{lee0}
Lee S., 2007, Mod. Phys. Lett. A. 22, 2003.

\bibitem[\protect\citeauthoryear{Livio and Stiavelli}{1998}]{livio}
Livio M. and Stiavelli M., 1998, Astrophys. J. 507, L13.

\bibitem[\protect\citeauthoryear{Luo and Su}{2005}]{luo}
Luo M. X. and Su Q. P., 2005, Phys. Lett. B. 626, 7.

\bibitem[\protect\citeauthoryear{Maeda}{1986}]{maeda}
Maeda K., 1986, Class. Quant. Grav. 3, 233.

\bibitem[\protect\citeauthoryear{Malec et al.}{2010}]{malec}
Malec A. L. et al., 2010, Mon. Not. Roy. Astron. Soc. 403, 1541.

\bibitem[\protect\citeauthoryear{Maor, Brustein and Steinhardt}{2001}]{maor1}
Maor I., Brustein R. and Steinhardt P. J., 2001, Phys. Rev. Lett. 86, 6.

\bibitem[\protect\citeauthoryear{Maor and Brustein}{2003}]{maor2}
Maor I. and Brustein R., 2003, Phys. Rev. D. 67, 103508.

\bibitem[\protect\citeauthoryear{Martin et al.}{2014}]{martin}
Martin J., Ringeval C., Trotta R. and Vennin V., 2014, JCAP 1403, 039. 

\bibitem[\protect\citeauthoryear{Masina and Notari}{2012}]{masina}
Masina and Notari, 2012, Phys. Rev. Lett., 108, 191302.

\bibitem[\protect\citeauthoryear{Mathiazhagan and Johri}{1984}]{mathia}
Mathiazhagan C. and Johri V. B., 1984, Class. Quantum Gravity, 1, L29.

\bibitem[\protect\citeauthoryear{Micheletti, Abdalla and Wang}{2009}]{micheletti}
Micheletti S., Abdalla E.and Wang B., 2009, Phys. Rev. D, 79, 123506.

\bibitem[\protect\citeauthoryear{Milne}{1937}]{milne}
Milne E. A., 1937, Proc. R. Soc. A3 242.

\bibitem[\protect\citeauthoryear{Mohamadnejad}{2019}]{ahmed}
Mohamadnejad A., 2019, Mod. Phys. Lett. A. 34(34), 1950277.

\bibitem[\protect\citeauthoryear{Moresco et. al.}{2012}]{ohd4}
Moresco M., Verde L., Pozzetti L., Jimenez R. and Cimatti A., 2012, J. Cosmol. Astropart. Phys 07, 053.

\bibitem[\protect\citeauthoryear{Mota and Barrow}{2004}]{mota}
Mota D. and Barrow J., 2004, MNRAS, 349, 291.

\bibitem[\protect\citeauthoryear{Murphy, Webb and Flambaum}{2003}]{murphy}
Murphy M. T., Webb J. K. and Flambaum V. V., 2003, Mon. Not. Roy. Astron. Soc. 345, 609.

\bibitem[\protect\citeauthoryear{Nordtvedt}{1970}]{nordtvedt}
Nordtvedt Jr K., 1970, ApJ, 161, 1059

\bibitem[\protect\citeauthoryear{Nunes and Lidsey}{2004}]{nunes}
Nunes N. J. and Lidsey J. E., 2004, Phys. Rev. D. 69, 123511.

\bibitem[\protect\citeauthoryear{Olive et. al.}{2002}]{olive0}
Olive K. A. et al., 2002, Phys. Rev. D. 66, 045022.

\bibitem[\protect\citeauthoryear{Olive and Pospelov}{2002}]{olive}
Olive K. A. and Pospelov M., 2002, Phys. Rev. D65, 085044.

\bibitem[\protect\citeauthoryear{Onofrio and Wegner}{2014}]{onofrio}
Onofrio R. and Wegner G. A., 2014, ApJ, 791, 125.

\bibitem[\protect\citeauthoryear{Padmanabhan and Roychoudhury}{2003}]{paddy}
Padmanabhan T. and Roychoudhury T., 2003, Mon. Not. R. Astron. Soc., 344, 823.

\bibitem[\protect\citeauthoryear{Pan, Mukherjee and Banerjee}{2018}]{supriyapan}
Pan S., Mukherjee A. and Banerjee N., 2018, MNRAS, 477, 1189.

\bibitem[\protect\citeauthoryear{Parkinson, Bassett and Barrow}{2004}]{parkinson}
Parkinson D., Bassett B. A. and Barrow J. D., 2004, Phys. Lett. B. 578, 235.

\bibitem[\protect\citeauthoryear{Rahmani et al.}{2013}]{rahmani}
Rahmani H. et al., 2013, Mon. Not. Roy. Astron. Soc. 435, 861.

\bibitem[\protect\citeauthoryear{Reinhold et al.}{2006}]{rien}
Reinhold E. et al., 2006, Phys. Rev. Lett. 96, 151101.

\bibitem[\protect\citeauthoryear{Riess}{2001}]{riess0}
Riess A., 2001, Astrophys. J., 560, 49.

\bibitem[\protect\citeauthoryear{Riess et al.}{2004}]{riess}
Riess A. G. et. al., 2004, Astrophys. J. 607, 665.

\bibitem[\protect\citeauthoryear{Riess et al.}{2018}]{riess2018}
Riess A. et al., 2018, ApJ, 855, 136.

\bibitem[\protect\citeauthoryear{Ryan et. al.}{2018}]{ryan}
Ryan J., Doshi S. and Ratra B.,2018, Mon. Not. Roy. Astron. Soc. 480, no. 1, 759.

\bibitem[\protect\citeauthoryear{Sahni et. al.}{2003}]{sahni}
Sahni V., Saini T., Starobinsky A. and Alam U., 2003, JETP Lett. 77 : 201 ; Pisma Zh. Eksp. Teor. Fiz 77 : 249.

\bibitem[\protect\citeauthoryear{Sandvik, Barrow and Magueijo}{2002}]{sandvik}
Sandvik H. B., Barrow J. D. and Magueijo J., 2002, Phys. Rev. Lett. 88, 031302.

\bibitem[\protect\citeauthoryear{Santos and Gregory}{1997}]{santos}
Santos C. and Gregory R., 1997, Ann. Phys., 258, 111.

\bibitem[\protect\citeauthoryear{Seikel, Clarkson and Smith}{2012}]{seikel}
Seikel M., Clarkson C. and Smith M., 2012, JCAP 1206, 036.

\bibitem[\protect\citeauthoryear{Sen, Sen and Sami}{2010}]{anjansen}
Sen S., Sen A. A. and Sami M., 2010, Phys. Lett. B. 686(1) 1.

\bibitem[\protect\citeauthoryear{Shafieloo, Kim and Linder}{2012}]{shafi}
Shafieloo A., Kim A. and Linder E., 2012, Phys. Rev. D 85, 123530.

\bibitem[\protect\citeauthoryear{Sikivie}{2008}]{sikivie}
Sikivie P., 2008, Lect. Notes Phys. 741, 19.

\bibitem[\protect\citeauthoryear{Simon, Verde and Jimenez}{2005}]{ohd1}
Simon J., Verde L. and Jimenez R., 2005, Phys. Rev. D. 71, 123001.

\bibitem[\protect\citeauthoryear{Slepian et. al.}{2014}]{slepian}
Slepian Z., Gott III, J. R., Zinn J., Mon. Not. R. Astron. Soc., 2014, 438(3), 1948.

\bibitem[\protect\citeauthoryear{Sol\`a et. al.}{2017}]{sola0}
Sol\`a J., Karimkhani E. and Khodam-Mohammadi A., 2017, Class. Quant. Grav. 34, no.2, 025006.

\bibitem[\protect\citeauthoryear{Sol\`a et. al.}{2019}]{sola3}
Sol\`a J., G\'omez-Valent A., Cruz P\'erez J. and Moreno-Pulido C., 2019, Astrophys. J. 886, no.1, L6.

\bibitem[\protect\citeauthoryear{Sol\`a et. al.}{2020}]{sola4}
Sol\`a J., G\'omez-Valent A., Cruz P\'erez J. and Moreno-Pulido C., 2020, arXiv:2006.04273v3 [astro-ph.CO].

\bibitem[\protect\citeauthoryear{Stern et. al.}{2010}]{ohd2}
Stern D., Jimenez R., Verde L., Kamionkowski M. and Stanford S. A., 2010, J. Cosmol. Astropart. Phys 02, 008.

\bibitem[\protect\citeauthoryear{Szydlowski et. al.}{2008}]{szyd}
Szydlowski M., Hrycyna O. and Kurek A., 2008, Phys. Rev. D. 77, 027302.

\bibitem[\protect\citeauthoryear{Thompson}{1975}]{thompson}
Thompson R. I., 1975, Astrophys. Lett. 16, 3.

\bibitem[\protect\citeauthoryear{Tsujikawa et al.}{2013}]{tsuji}
Tsujikawa S., Ohashi J., Kuroyanagi S. and De Felice A., 2013, Phys. Rev. D., 88, 023529.

\bibitem[\protect\citeauthoryear{Ubachs et al.}{2016}]{ubachs}
Ubachs W., Bagdonaite J., Salumbides E. J., Murphy M. T. and Kaper L., 2016, Rev. Mod. Phys. 88, 021003.

\bibitem[\protect\citeauthoryear{Upadhye, Ishak and Steinhardt}{2005}]{upadhye}
Upadhye A., Ishak M. and Steinhardt P. J., 2005, Phys. Rev. D. 72, 063501.

\bibitem[\protect\citeauthoryear{Uzan}{2003}]{uzan1}
Uzan J. P., 2003, Rev. Mod. Phys. 75, 403.

\bibitem[\protect\citeauthoryear{Uzan}{2011}]{uzan2}
Uzan J. P., 2011, Living Rev. Rel. 14, 2.

\bibitem[\protect\citeauthoryear{Wagoner}{1970}]{wagoner}
Wagoner R. V., 1970, Phys. Rev. D, 1, 3209.

\bibitem[\protect\citeauthoryear{Webb et. al.}{2001}]{webb}
Webb J. K. et al., 2001, Phys. Rev. Lett. 87, 091301.

\bibitem[\protect\citeauthoryear{Weerdenburg et al.}{2011}]{weerd}
Weerdenburg F., Murphy  M. T., Malec A. L., Kaper L. and Ubachs W., 2012, Phys. Rev. Lett. 106, 180802.

\bibitem[\protect\citeauthoryear{Wendt and Molaro}{2011}]{wendt}
Wendt M. and Molaro P., 2011, Astron. Astrophys. 526, A96.

\bibitem[\protect\citeauthoryear{Wetterich}{1988}]{wetterich}
Wetterich C., 1988, Nucl. Phys. B302, 645 ; Nucl. Phys. B302, 668.

\bibitem[\protect\citeauthoryear{Yang et al.}{2018}]{yang}
Yang Y. B., Liang J., Bi Y. J., Chen Y., Draper T., Liu  K. F. et al., 2018, Phys. Rev. Lett. 121, 212001.

\bibitem[\protect\citeauthoryear{Zhang et. al.}{2014}]{ohd6}
Zhang C., Zhang H., Yuan S., Zhang T. J. and Sun Y. C., 2014, Res. Astron. Astrophys. 14, 1221.





















\end{thebibliography}

\end{document}